\documentclass[usenatbib]{mnras}
\usepackage[utf8]{inputenc}
\usepackage{graphicx}
\usepackage{xcolor}
\usepackage{multirow}
\usepackage{float}

\newcommand{\HII}{H\,\textsc{ii}}
\newcommand{\NeII}{\textrm{Ne\,{\textsc{ii}}}}
\newcommand{\NeIII}{\textrm{Ne\,{\textsc{iii}}}}
\newcommand{\SIV}{\textrm{S\,{\textsc{iv}}}}

\title{The spatially resolved PAH characteristics in the Whirlpool Galaxy (M51a)}

\author[R.X. Zang et al.]{Rong Xuan Zang,$^{1,2}$\thanks{E-mail: rzang@uwo.ca} Alexandros Maragkoudakis,$^{1,3,6}$ Els Peeters$^{1,4,5}$\\
$^{1}$Department of Physics \& Astronomy, University of Western Ontario, London, ON, N6A 3K7, Canada\\
$^{2}$Department of Microbiology \& Immunology, University of Western Ontario, London, ON, N6A 3K7, Canada\\
$^{3}$NASA Ames Research Center, MS 245-6, Moffett Field, CA 94035-1000, USA \\
$^{4}$Institute for Earth and Space Exploration, University of Western Ontario, London, ON, N6A 3K7, Canada\\
$^{5}$SETI Institute, 189 Bernardo Avenue, Suite 100, Mountain View, CA 94044, USA \\
$^{6}$Universities Space Research Association, Columbia, MD, USA}

\date{December 2021}

\begin{document}

\maketitle

\begin{abstract}
    We present a detailed study on the spatially resolved polycyclic aromatic hydrocarbon (PAH) emission properties in the (circum)nuclear region (NR) and extranuclear regions (ENRs) of M51a using \textit{Spitzer}-IRS observations. Correlations among PAH intensity ratios are examined with respect to each other, local physical parameters, galactocentric distance ($R_{g}$), and very small grain (VSG) emission. Additional comparison is performed with the mid-infrared emission features in the \HII{} regions of M33 and M83. 
    The NR exhibits the strongest correlation among the PAH intensity ratios, whereas ENRs are showing increased scatter attributed to ISM emission. Overall, the radiation field hardness has a higher impact on PAH emission than metallicity, with the latter regulating PAH variance as a function of $R_{g}$. Specifically, the variance of PAH emission with respect to the different physical parameters suggests a higher rate of small/medium PAH processing compared to large PAHs and a higher ratio of small-to-large PAHs formed with increasing galactocentric distance. We find similarities between the 7.7 \micron{} carriers in M51a's NR and M83's \HII{} regions, the 8.6 \micron{} carriers in M51a's NR and M33 \HII{} regions, and both types of carriers between M51a's ENRs, M33's, and M83's \HII{} regions.
    We have identified a positive correlation between PAH/VSG and the PAH intensity ratios. We conclude that the relative abundance of PAHs and VSG is not solely driven by the hardness of the radiation field. 
\end{abstract}


\section{Introduction}
\label{sec:introduction}
The mid-infrared (IR) spectra of a large fraction of astronomical sources are dominated by emission at 3.3, 6.2, 7.7, 8.6, and 11.3 $\mu$m, typically attributed to polycyclic aromatic hydrocarbons \citep[PAHs;][]{Leger1984, Allamandola1985}. PAHs are principally carbon atoms fused together in benzene rings with hydrogen attached to the edges. PAHs absorb ultra-violet (UV) photons which is followed by IR fluorescence via their vibrational modes \citep{Rossi1983,Kiefer1985,Allamandola_1989}. PAH emission features have been observed across a variety of sources from reflection nebulae, photodissociation region, and galaxies \citep[e.g.][]{Hony2001, Peeters2002, Smith07b, Galliano2008}. Taken together, PAHs accounts for $\sim 10-30\%$ of the interstellar carbon \citep{Snow1995} and contribute up to 20\% of the total IR emission in galaxies \citep{Smith07b}, constituting a significant carrier of the observed IR emission of galactic and extragalactic sources.

The characteristics of the PAH emission varies among different sources or even within a given source \citep[e.g.][]{Peeters2002, BregmanTemi2005, Berne2007, Smith07b, Boersma_2016, Peeters2017, Maragkoudakis2018b}. This variation is due to the range of intrinsic properties of the PAHs (such as charge, size, and molecular structure) making up the PAH population, which in turn is influenced by the local physical conditions (i.e. radiation field, density, temperature) of the environments in which they reside \citep[e.g.][]{Allamandola1999, Schutte1993, Sloan2007, Galliano2008, Ricca2012, Stock2017, Maragkoudakis2020}. For instance, emission in the 6-9 $\mu$m range dramatically increases in comparison with the 3 $\mu$m and 10-15 $\mu$m range for a given PAH molecule upon ionization \citep{Allamandola1999}. Therefore, ratios of the 6-9 $\mu$m PAH emission to the 3.3 or 11.2 $\mu$m bands are employed to determine the PAH charge balance. Likewise, the ratio of the total PAH emission to the very small grains (VSG) emission also varies among sources and within extended sources with increasing hardness of the radiation field, indicative of PAH processing or destruction \citep[e.g.][]{Madden2006, Smith07b, Maragkoudakis2018b}. In nearby galaxies, the variance in PAH emission is also evident within different galactic environments and radiation fields. In galaxies of low-metallicity the PAH intensities are typically weakened \citep[e.g.][]{Engelbracht2005,Sandstrom2012}, but there is also evidence of a suppression in PAH strengths of active galactic nuclei (AGN) hosts \citep[][]{Smith07b}.

With the growing number of spatially resolved galactic surveys, the understanding and characterization of PAH emission and their variation in resolved galactic regions is essential. Following the work presented in \citet{Maragkoudakis2018b}, where the authors examined the PAH emission properties in resolved \HII{} regions in the nearby star-forming galaxies M33 and M83, we expand the spatially resolved characterization of PAH emission in the Whirlpool galaxy (also known as M51a or NGC~5194). In addition, we examine M51a's circumnuclear region to characterize and compare the nuclear against extranuclear PAH emission.

This paper is organized as follows: Section \ref{target} gives a brief description of M51a and Section \ref{sec:Methodology} presents the observations, data reduction, and data analysis details. The results are presented and discussed in Section \ref{sec:results_discussion}, and a summary of our conclusions is presented in Section \ref{sec:Summary_conclusion}.

\section{Whirlpool Galaxy}
\label{target}

M51a is a nearby (at a distance of 7.1 Mpc; \citealt{Takats2006}) almost face-on, grand-design spiral galaxy currently undergoing a merger with its smaller neighbor M51b (NGC~5195). Optical nuclear activity diagnostics classify the nucleus of M51a as either a Low Ionization Nuclear Emission Region \citep{Carrillo1999, Satyapal2004} or Seyfert 2 \citep{Ho1997b}, with the Seyfert 2 activity potentially triggered as a result of the interaction with its companion \citep[e.g.][]{Koulouridis2014}. The average oxygen abundance of M51a is $12 + \mathrm{log(O/H)}$ = 8.54 with a -0.31 dex $\rho_{25}^{-1}$\footnote{$\rho_{25}^{-1}\equiv$ radius of the major axis at the $\mu_{B}$ = 25 mag arcsec$^{-2}$ isophote \citep{deV1991rc3}.} slope of the radial abundance gradient \citep{Moustakas2010}. In addition, the numerous \HII{} regions present in M51a are reported to have roughly solar metallicities \citep{Bresolin2004, Moustakas2010, Croxall2015} although, the derived measurements depend on the adopted method (theoretical or empirical) used to determine the oxygen abundances. Its low inclination, which enables a clean and free from geometric complication study, prominent spiral structure, which allows the determination of pure star-forming regions, and wealth of \textit{Spitzer}-IRS archival observations constitutes M51a as an optimal source for a detailed study of the spatially resolved PAH emission characteristics in the local Universe.

Recently, \cite{Zhang2021} also examined M51 using a different method of extracting PAH emission information from \textit{Spitzer}-IRS spectra with only partial coverage, and investigated the effectiveness of PAH emission as a star-formation rate (SFR) indicator on sub-kpc scales, where they concluded that PAHs serve as excellent tracers of the SFR in a wide range of environments, from $\sim 0.4$ kpc close to the nucleus to 6 kpc out in the disk of the galaxy.

\section{Observations and Analysis}
\label{sec:Methodology}

\subsection{Mid-IR observations}
\label{subsec:MidIRObservations}

\begin{figure}
        \begin{center}
        \includegraphics[keepaspectratio=true,width=.45\textwidth]{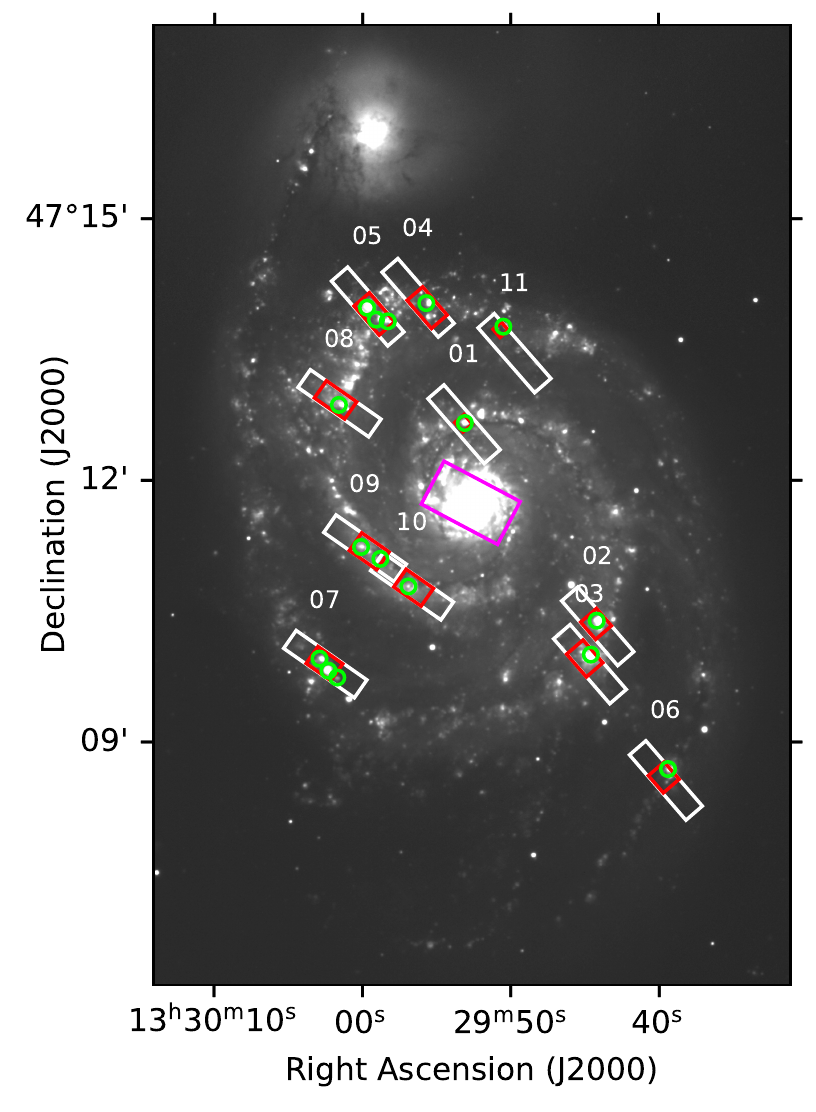}
\caption{H$\alpha$ image of M51a (obtained from the Spitzer
Infrared Nearby Galaxies Survey--SINGS; \citealt{Kennicutt2003}). Nuclear and extranuclear \textit{Spitzer}-IRS apertures are shown in purple and white colors respectively, star-forming apertures are shown in red (see Section \ref{subsec:sfvsism}), and H~\textsc{ii} regions mapped by \protect\cite{Croxall2015} falling within the IRS apertures are shown in green circles.}
       \label{fig:Hamap}
        \end{center}
\end{figure}

M51a \textit{Spitzer} Infrared Spectrograph \citep[IRS;][]{Werner, Houck} 3D spectral cube observations were retrieved from the SINGS \citep{Smith07b} Spitzer Legacy IRSA database\footnote{\href{https://irsa.ipac.caltech.edu/data/SPITZER/SINGS/}{https://irsa.ipac.caltech.edu/data/SPITZER/SINGS/}}. This dataset includes short-low (SL1 and SL2) mapping observations from the circumnuclear region (referenced as nuclear region or NR hereafter) of the galaxy and 11 extra-nuclear regions (referenced as ENRs hereafter)\footnote{The raw spectral cubes use a region numbering convention starting from ``00". Here, we adopt a region numbering starting from ``01" for convenience.} covering the 5--15 $\mu$m wavelength range. The spectral cubes were pre-assembled, as part of the SINGS Data Release 5 (DR5), using \textsc{cubism} \citep{Smith07a} with background subtraction and flux calibration performed. The ENR observations target star-forming clumps along the galaxy's spiral arms at different galactocentric radii (Fig. \ref{fig:Hamap}). The ENR targets were optically selected, and therefore in certain cases a mismatch between the peak of the H$\alpha$ emission and 8 \micron{} emission maps may be observed.

\begin{table*}
        \begin{center}
        \caption{PAH fluxes and equivalent widths (EQW) for the M51a NR and ENRs. PAH fluxes are in units of $10^{-16}W/m^2$ and EQWs are in $\mu$m.}
        \label{tab:M51flux}
        \begin{tabular}{@{}lcccccccccc}    
        \hline
        Regions&\multicolumn{2}{c}{6.2 $\mu$m PAH}&\multicolumn{2}{c}{7.7 $\mu$m PAH}&\multicolumn{2}{c}{8.6 $\mu$m PAH}&\multicolumn{2}{c}{11.2 $\mu$m PAH}&\multicolumn{2}{c}{12.6 $\mu$m PAH}\\
&Flux&EQW&Flux&EQW&Flux&EQW&Flux&EQW&Flux&EQW\\
        \hline
        NR&5.05$\pm$0.27&1.58&19.34$\pm$1.07&5.53&3.37$\pm$0.22&0.97&5.61$\pm$0.13&1.86&3.46$\pm$0.16&1.28\\
        1&2.12$\pm$0.12&3.40&5.60$\pm$0.30&7.36&1.03$\pm$0.06&1.37&1.44$\pm$0.03&2.39&0.89$\pm$0.04&1.58\\
        2&2.88$\pm$0.11&9.02&7.19$\pm$0.14&11.42&1.58$\pm$0.06&2.15&1.79$\pm$0.03&2.21&1.00$\pm$0.05&1.24\\
        3&2.50$\pm$0.08&13.69&7.56$\pm$0.27&16.91&1.40$\pm$0.06&2.83&1.56$\pm$0.03&2.47&0.86$\pm$0.05&1.42\\
        4&2.00$\pm$0.07&10.43&6.88$\pm$0.22&21.02&1.28$\pm$0.05&5.74&1.47$\pm$0.03&3.93&0.78$\pm$0.05&3.07\\
        5&2.18$\pm$0.07&13.43&7.30$\pm$0.23&23.09&1.39$\pm$0.05&7.63&1.49$\pm$0.03&5.52&0.79$\pm$0.05&2.96\\
        6&1.48$\pm$0.07&16.57&4.71$\pm$0.24&32.86&0.86$\pm$0.05&9.82&0.89$\pm$0.03&3.76&0.45$\pm$0.05&2.26\\
        7&1.59$\pm$0.11&5.58&4.33$\pm$0.21&11.92&1.08$\pm$0.07&6.77&1.07$\pm$0.04&3.90&0.51$\pm$0.06&1.59\\
        8&2.70$\pm$0.13&3.78&7.89$\pm$0.30&8.16&1.75$\pm$0.08&1.87&1.95$\pm$0.04&2.10&1.02$\pm$0.06&1.04\\
        9&2.35$\pm$0.12&2.04&7.72$\pm$0.34&6.29&1.71$\pm$0.08&1.47&2.07$\pm$0.04&1.99&1.10$\pm$0.06&1.15\\
        10&2.71$\pm$0.12&2.24&8.80$\pm$0.38&6.47&1.95$\pm$0.08&1.53&2.33$\pm$0.04&2.09&1.23$\pm$0.07&1.15\\
        11&1.52$\pm$0.09&5.68&5.42$\pm$0.31&11.48&0.92$\pm$0.06&1.60&1.24$\pm$0.03&2.06&0.71$\pm$0.05&1.2\\
        \hline
        \end{tabular}
        \end{center}
\end{table*}

\subsection{Spectral extraction and stitching}
\label{subsec:dataReduction}

In order to spatially align and combine the SL1 and SL2 cubes, we re-gridded the SL2 cubes based on the SL1 astrometries. Subsequently, to achieve better optimization in terms of signal-to-noise (S/N) ratio for the spatially resolved mid-IR feature examination, we performed a $2''\times2''$ binning among adjacent pixels. Furthermore, we determined corrections to match the SL2 and SL1 orders in their overlap region. However, since an application of relatively large or small scaling factors to the SL2 spectra can artificially enhance or decrease the 6.2 $\mu$m PAH feature, we adopt a conservative scale factor range between 0.8 and 1.2 rejecting pixels with corrections factors outside this range. This particular boundary choice of scale factors is consistent with the absolute flux calibration of IRS (20\%). The re-gridded, binned and scaled (in the case of SL2 mode) spectra were combined to create the final NR and ENR spectral cubes. We extracted spectra from each pixel in the final cubes, as well as the integrated spectrum from all the pixels in each region (NR and ENRs), where the size of the low-resolution aperture used to extract the nuclear region spectrum was $30''\times52''$, while the ENR spectra were extracted from $15''\times52''$ apertures.

\subsection{Spectral decomposition and PAH band emission measurement}
\label{subsec:PAHFIT}

\begin{figure}
    \centering
    \hspace*{-0.8cm}\includegraphics[scale=0.45]{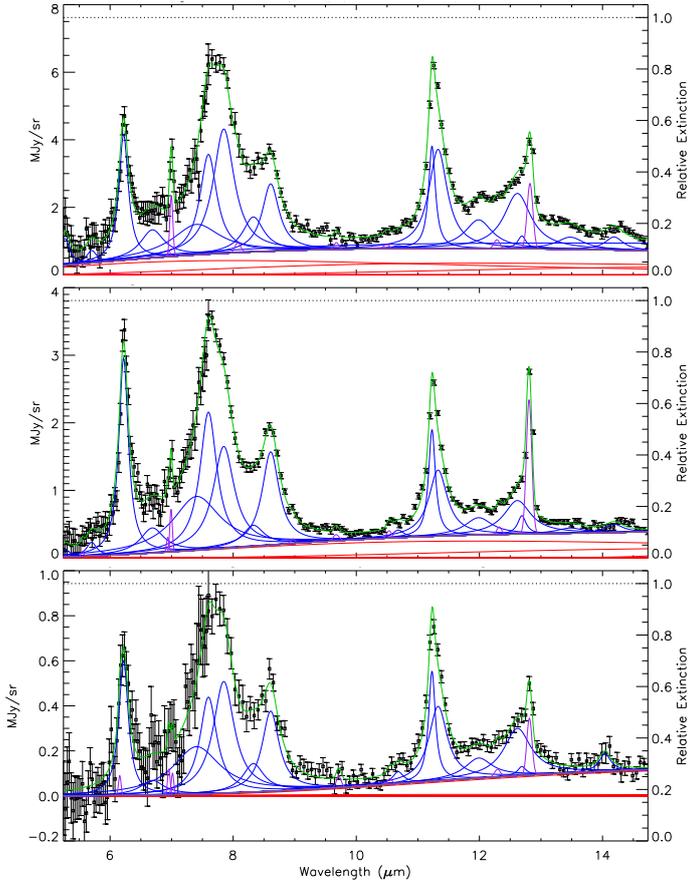}
    \caption{Example spectra and \textsc{pahfit} decomposition for an NR spectrum (top panel), an ENR SF spectrum (middle panel), and an ENR ISM spectrum (bottom panel). See Section \ref{sec:sfismreg} for the ENR SF and ISM definition.}
    \label{fig:example_spec}
\end{figure}

We used the \textsc{pahfit} \citep{Smith07b} spectral decomposition code to analyze the combined $2''\times2''$ binned NR and ENR spectra. \textsc{pahfit} fits the input spectrum using a combination of different components that consist of the starlight continuum, thermal dust continua, $H_2$ rotational lines, fine-structure lines, and dust emission features, while further taking into account dust extinction. Dust extinction due to strong silicate absorption features at 9.7 and 18 \micron{} can suppress PAH emission primarily in the 8.6 and 11.2 \micron{} bands, and secondarily in the 7.7 \micron{} complex region. \textsc{pahfit} handles the extinction offering a set of extinction curves with fully mixed dust geometries or uniform foreground screens of dust. However, the inclusion of extinction as a free parameter in the fit can potentially lead to degenerate solutions and thus high uncertainty in the evaluated strengths of PAH features. By allowing extinction to freely vary on \textsc{pahfit} we found that silicate absorption is negligible ($\tau_{9.7} \leq 0.1$) for 98.5\% of the sample. For the pixels that were fitted with extinction values higher than 0.5 we examined the fits of the surrounding pixels. In all cases the neighboring pixels had negligible extinction compared to the pixel of inspection. Subsequently, following the assumption of a uniform distribution of dust attenuation and extinction within the galaxy at small scales, we repeated the analysis setting now manually the extinction to zero for all pixels. Fig. \ref{fig:example_spec} presents example spectra and \textsc{pahfit} decomposition for the NR and ENRs. Table \ref{tab:M51flux} presents the measured feature strengths and equivalent widths of the prominent PAH bands for the integrated spectra of the regions. 

\subsection{Star-forming and ISM regions} \label{sec:sfismreg}
\label{subsec:SFR_ISM}

In order to investigate the relative contribution to the PAH emission from regions dominated by star-forming (SF) activity as opposed to the Interstellar Medium (ISM), we defined squared star-forming apertures within the ENR IRS observations that: i) contain the brightest star-forming knots in the H$\alpha$ map (Fig. \ref{fig:Hamap}); ii) encompass the bulk emission from the H\,\textsc{ii} regions mapped by \citet[][hereafter C15]{Croxall2015} excluding ISM contribution. Pixels lying outside the SF apertures were labeled as ISM regions. Note though that while the previous criteria define the regions of most intense star formation, ISM defined regions might include portion of SF contribution, as their pixels sample a part of the galaxy's spiral arms.

\begin{figure*}
\includegraphics[width=\textwidth]{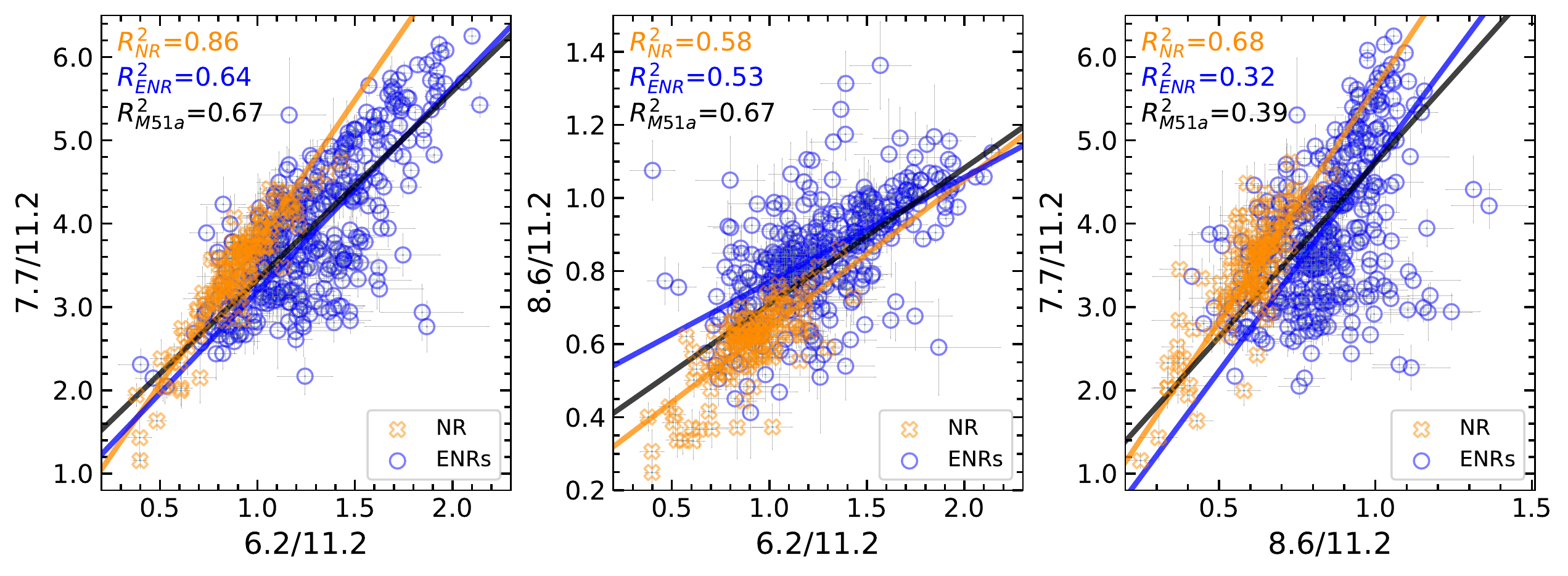}
\caption{Relative comparison among the 7.7/11.2, 6.2/11.2, and 8.6/11.2 PAH intensity ratios of M51a's NR (orange crosses) and ENRs (blue circles). Linear regression fits and regression coefficients ($R^{2}$) for the NR and ENRs are shown in orange and blue respectively, while the fit and $R^{2}$ of all regions is shown in black. The slopes and intercept statistics of the linear regression fit are given in Table \ref{Apptab:M51slopePAHRatios}.}
\label{fig:M51-Overview}
\end{figure*}

\subsection{Properties of the star-forming regions}
\label{subsec:OpticalObservations}

For a comparison between the mid-IR emission characteristics and the physical properties of the \HII{} regions they are associated with, we used the \HII{} observations from C15. We gathered their deprojected galactocentric radii (in units of the isophotal radius $R_{25}$\footnote{The radius where a galaxy's surface brightness falls to a level of 25 B-mag/arcsec$^{2}$}; $R_{g}/R_{25}$), electron temperatures ($T_{e}$), electron densities ($n_{e}$), and oxygen abundances ($12 + \mathrm{log(O/H)}$). The association of the SL ENR apertures with the HII{} regions observations in C15 was performed visually by over-plotting the \HII{} region apertures on top of the IRS apertures (Fig. \ref{fig:Hamap}). While we obtained $R_{g}/R_{25}$ for all ENRs of M51a, no $T_{e}$, $n_{e}$, or oxygen abundances measurements were available for ENR 1 and ENR 10. Therefore, only the remaining 9 ENRs were used for our analyses involving $T_{e}$, $n_{e}$, or oxygen abundances.

C15 derived oxygen abundances using a mixture of direct and semi-empirical methods. The direct method involves the measurement of temperature-sensitive auroral lines such as [N\,\textsc{ii}]\,$\lambda5755$ and [S\,\textsc{iii}]\,$\lambda6312$, while in the semi-empirical method the relative N/O abundance ratio was calculated assuming an electron temperature consistent with an adopted strong-line oxygen abundance. The electron densities were calculated from the [S\,\textsc{ii}]\,$\lambda\lambda6717,6731$ line ratio, while electron temperatures were derived from scaling relations involving the [O\,\textsc{ii}], [O\,\textsc{iii}], [N\,\textsc{ii}], and [S\,\textsc{iii}] temperature-sensitive lines.

\section{Results and discussion}
\label{sec:results_discussion}

We examined the PAH emission characteristics in the (circum)nuclear region and 11 extranuclear regions of M51a as a function of local physical parameters (hardness of radiation field [S\,\textsc{iv}]/[Ne\,\textsc{ii}], oxygen abundance ($12 + \mathrm{log(O/H)}$), $T_{e}$, $n_{e}$) and galactocentric radii, further assessing the characteristics of the relative PAH emission ratio correlations (Section~\ref{subsec:pahratios}). We also investigated the total PAH to very small grain (VSG) emission with respect to: (i) the hardness of the radiation field as probed through the [S\,\textsc{iv}]/[Ne\,\textsc{ii}] emission ratio and (ii) the ionization fraction as traced by the three PAH intensity ratio (Section~\ref{subsec:PAH_VSGData}). 
A comparison with the PAH emission and total PAH to VSG emission in the H\,\textsc{ii} regions of star-forming galaxies M33 and M83 is made. Throughout the analysis, we employed the coefficient of determination ($R^{2}$) to characterize scatter among fitted parameters and the weighted Pearson correlation coefficient ($p_{w}$) to asses the linearity of a correlation.

\subsection{Relative PAH intensities} 
\label{subsec:pahratios}

\begin{figure*}
\includegraphics[width=\textwidth]{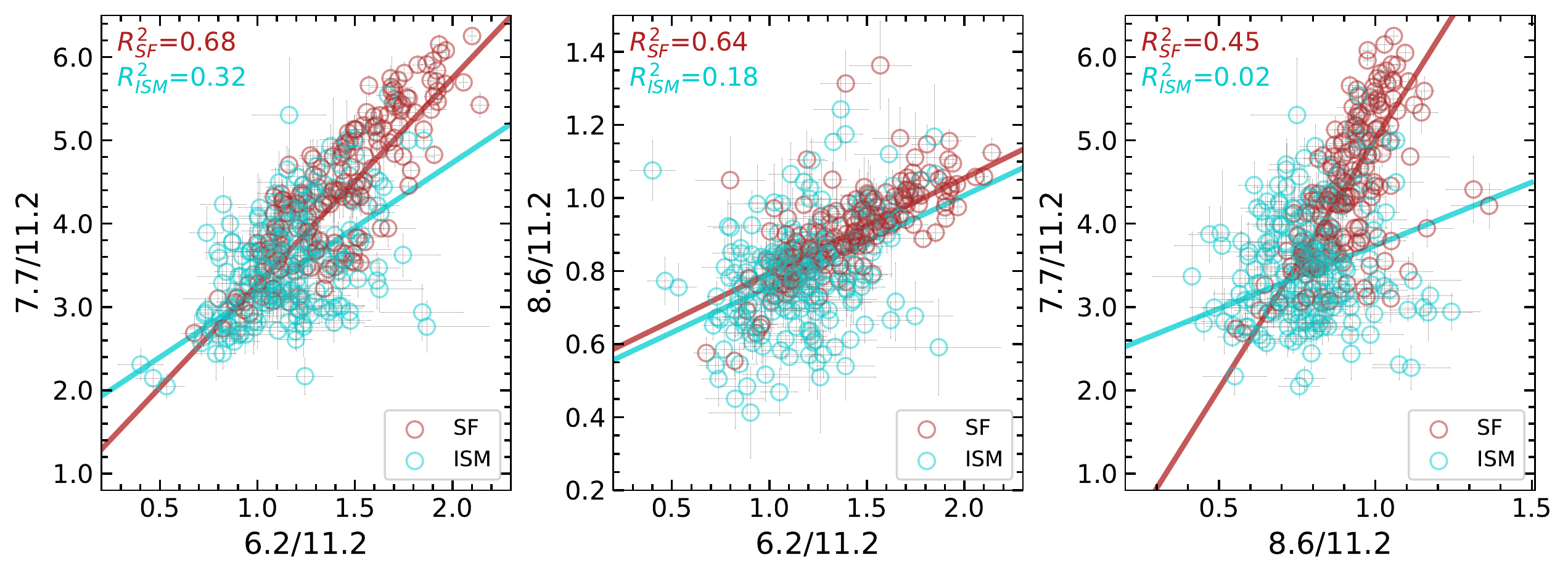}
\caption{Relative comparison among the 7.7/11.2. 6.2/11.2, and 8.6/11.2 PAH intensity ratios for star-formation dominated (red) and ISM dominated (cyan) pixels within the M51a ENRs. The corresponding linear regression fit and regression coefficient ($R^2$) values are displayed.}
\label{fig:M51SFvsISM}
\end{figure*}

\begin{table*}
    \begin{center}
    \caption {Weighted Pearson coefficients ($p_{w}$) of the various PAH intensity correlations in M51a, M33, and M83.} 
    \label{tab:M51pearson}
    \begin{tabular}{@{}lcccccccc}
    
    \hline
    Regions&7.7/11.2 &8.6/11.2  &7.7/11.2 &log(PAH/VSG) &PAH/VSG &PAH/VSG &PAH/VSG \\ 
    &vs       &vs        &vs       &vs  &vs  &vs  &vs  \\
    &6.2/11.2 &6.2/11.2  &8.6/11.2 &log([S\,\textsc{iv}]/[Ne\,\textsc{ii}])&6.2/11.2 &7.7/11.2 &8.6/11.2\\
    \hline
    NR &0.92&0.73&0.79&-0.23&0.78&0.76&0.52\\ 
    ENRs (all)  &0.83&0.73&0.62&-0.11&0.20&0.02&0.12\\ 
    ENR 1       &0.85&0.64&0.84&-1.00&0.56&0.41&0.28\\ 
    ENR 2       &0.89&0.70&0.79&-0.55&0.44&0.22&0.18\\ 
    ENR 3       &0.90&0.87&0.82&0.46&-0.04&-0.16&-0.18\\ 
    ENR 4       &0.94&0.86&0.80&-0.95&0.20&-0.06&0.05\\ 
    ENR 5       &0.96&0.74&0.80&-0.75&-0.10&-0.21&-0.42\\ 
    ENR 6       &0.82&0.69&0.39&-1.00&0.17&0.18&0.26\\ 
    ENR 7       &0.85&0.12&-0.03&1.00&-0.22&-0.31&0.30\\ 
    ENR 8       &0.95&0.76&0.65&0.63&0.21&0.18&0.26\\ 
    ENR 9       &0.91&0.84&0.68&1.00&0.55&0.68&0.44\\ 
    ENR 10      &0.92&0.76&0.73&-0.15&0.70&0.81&0.42\\ 
    ENR 11      &0.93&0.83&0.85&0.98&0.75&0.70&0.55\\ 
    ENRs SF     &0.84&0.79&0.68&\ldots&\ldots&\ldots&\ldots\\ 
    ENRs ISM    &0.59&0.46&0.23&\ldots&\ldots&\ldots&\ldots\\ 
    ENRs 1-2, 9-11   &\ldots&\ldots&\ldots&\ldots&0.57&0.51&0.37\\ 
    M51a        &0.85&0.82&0.67&\ldots&0.39&0.05&0.33\\ 
    M33&0.94&0.94&0.86&-0.51&0.62&0.67&0.50\\ 
    M83&0.96&0.80&0.85&-0.64&0.47&0.37&-0.003\\
    \hline
\end{tabular}
\end{center}
\end{table*}

Fig. \ref{fig:M51-Overview} presents the--previously established--relative PAH intensity correlations among the 6.2/11.2, 7.7/11.2, and 8.6/11.2 ratios for M51a's NR and ENRs. The slopes and intercept statistics of the linear regression fits can be found in Table \ref{Apptab:M51slopePAHRatios}. These ratios trace the PAH charge balance of the emitting PAH population \citep{Galliano2008, Boersma_2016}. 
Overall, M51a NR presents the most robust correlations compared to M51a ENRs with coefficients of determination $R^{2}_{NR} \geq 1.1\times R^{2}_{ENR}$ and weighted Pearson's $p_{w NR} \geq p_{w ENR}$ (Table \ref{tab:M51pearson}). The 7.7/11.2 -- 6.2/11.2 intensity ratios are best correlated, among the different PAH intensity ratios considered here, in both the NR and ENRs, while the 8.6/11.2 -- 6.2/11.2 ratios have the weakest regression relationship in the NR, with ENRs moderately correlating (Fig. \ref{fig:M51-Overview}; Table \ref{tab:M51pearson}). The 7.7/11.2 -- 8.6/11.2 intensity ratios have the weakest regression relationship for the ENRs, however the NR has the second firmest correlation, following 7.7/11.2 -- 6.2/11.2.
Furthermore, examination of the individual ENRs reveals moderate to strong correlations between the relative PAH intensity ratios with only two ENRs (6 and 7) presenting weak correlations (Table \ref{tab:M51pearson}). 

Considering all NR and ENRs combined, the correlation coefficients of M51a are $p_{w}=0.85$, $p_{w}=0.82$, and $p_{w}=0.67$ for the 7.7/11.2 -- 6.2/11.2, 8.6/11.2 -- 6.2/11.2, and 7.7/11.2 -- 8.6/11.2 correlations respectively. Based on ISOCAM-CVF observations and employing two spectral decomposition methods for the aromatic features, i.e. spline ($S$) and Lorentzian ($L$), \cite{Galliano2008} derived for M51 correlation coefficients of $p_{w(S)}=0.62$ and $p_{w(L)}=0.65$ for the 7.7/11.2 -- 6.2/11.2 correlation, and $p_{w(S)}=0.41$ and $p_{w(L)}=0.65$ for the 7.7/11.2 -- 8.6/11.2 correlation. Although there is a good agreement between the respective $p_{w}$ values in this work and in \cite{Galliano2008}, deviations are mostly due to differences in the observations and spectral resolution of $Spitzer$-IRS and ISOCAM-CVF, as well as in the analysis methods used to measure the aromatic features, with Lorentzian profiles producing results comparable to the Drude profiles adopted in \textsc{pahfit}.

Among the three PAH intensities ($I_{6.2}$, $I_{7.7}$, $I_{8.6}$) involved in the previous correlations, the 8.6 \micron{} feature is typically most affected by the presence of a silicon absorption feature at 10 \micron{}.
As a consequence, the proper modeling and recovery of the 8.6 \micron{} feature profile, and therefore its intensity, can often be challenging and/or subjected to larger uncertainties. Typically, the increased scatter observed in correlations involving the 8.6 \micron{} PAH band has been attributed to extinction and to its lower intensity, compared to that of the 6.2 and 7.7 \micron{} PAH bands. However, we found extinction to be negligible in most pixels of our NR and ENR maps (Section~\ref{subsec:PAHFIT}). In addition, recent results on the reflection nebula NGC~2023 \citep{Peeters2017, Sidhu2021} indicate that the ionic bands at 6.2, 7.7, and 8.6 \micron{} have distinct spatial morphologies and characteristics; we are thus probing differences in the underlying PAH population contributing to these bands. As a consequence, the differences in degree of correlation between them, as reported in the literature and in this work, are in part due to the different subclass of PAH populations responsible for their emission rather than solely due to extinction effects.

\subsubsection{Star-forming and ISM region emission} 
\label{subsec:sfvsism}

The observed PAH emission arises from vibrational de-excitation after the absorption of a photon. The photon absorption efficiency of PAHs is largest for UV photons which also carry more energy compared to e.g. visual photons. As such, when UV radiation is present such as in \HII{} regions, UV photons drive the IR PAH emission. We examined the emission from star-forming (SF) and non-SF ISM pixels (referred to as ISM pixels hereafter) separately (Fig. \ref{fig:M51SFvsISM}) to further assess any additional trends in the PAH intensity ratio correlations. Overall the SF pixels present the strongest correlations, while the ISM pixels in the 8.6/11.2 -- 6.2/11.2 and 7.7/11.2 -- 8.6/11.2 plots lack any correlation ($R_{ISM}^2=0.18, p_{w ISM}=0.46$ and $R_{ISM}^2=0.02, p_{w ISM}=0.23$ respectively), despite some overlap with the SF pixels. This illustrates that the PAH emission from the ISM region is the main contributor to the observed scatter in the intensity ratio correlations described for all pixels (Fig. \ref{fig:M51-Overview}). Further examination of the S/N ratio of the ISM and SF pixels, showed that both cases extend to similarly low values, and therefore it is not the S/N of the ISM regions that drive the scatter. As a result, this may imply a difference in the dominant PAH populations within these regions as also reported by \citet{Stock2017}, \citet{knight_2020} and \citet{Knight:orion}.


\subsubsection{Dependence on the physical conditions}
\label{sec:PAH_ratios-Phys_cond}

\begin{figure*}
\includegraphics[width=\textwidth]{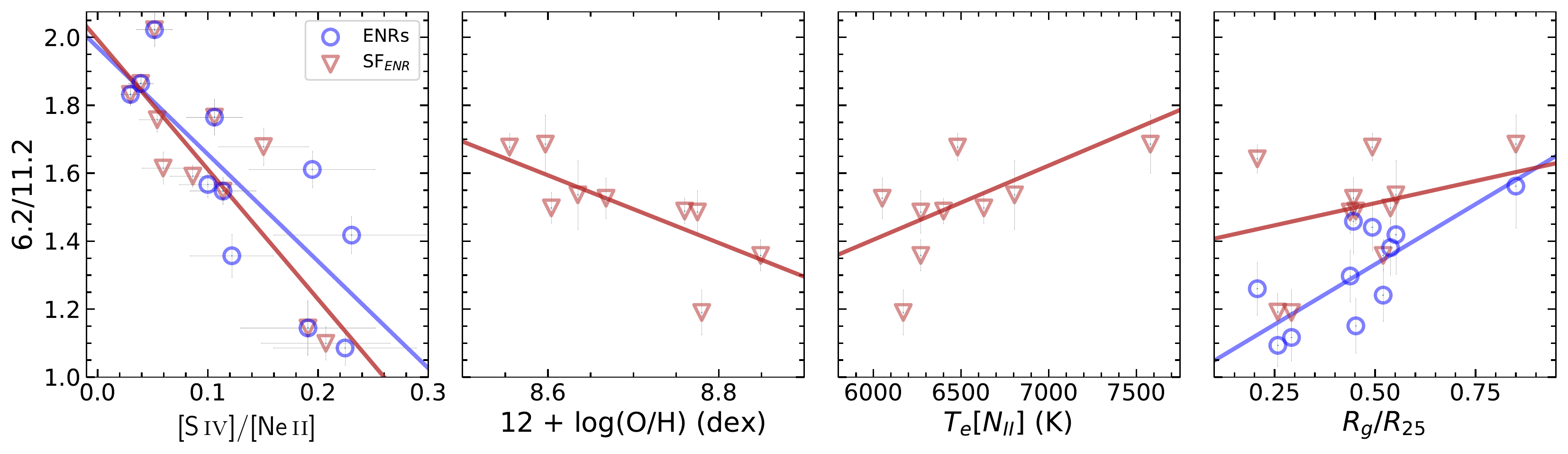}
\caption{Average 6.2/11.2 PAH intensity ratios of M51a SF regions and entire ENRs, as a function of local physical parameters. From left to right column: hardness of the radiation field as probed by} [\SIV/\NeII], oxygen abundance ($12 + \mathrm{log(O/H)}$), electron temperature ($T_e$) as determined by \protect\cite{Croxall2015}, and deprojected galactocentric distance ($R_g/R_{25}$) are displayed. Note for oxygen abundance and electron temperature only M51a SF regions are plotted because they are derived from IR lines present only in those regions. Weighted Pearson's ($p_w$) coefficients and linear regression fits and regression coefficients ($R^2$) for all PAH intensity ratios are displayed in Table \ref{apptab:M51fluxPropertiesRandp}, with the corresponding linear parameters provided in Table \ref{apptab:phys_prop_slopes}.
\label{fig:M51PAHPhysicalProperties}
\end{figure*}

We examined the dependence of the relative PAH intensity ratios on their local physical conditions for the integrated emission of each ENR. Fig. \ref{fig:M51PAHPhysicalProperties} presents the relation of the 6.2/11.2 PAH intensity ratio with hardness of the radiation field ([\SIV/\NeII]), oxygen abundance ($12 + \mathrm{log(O/H)}$), electron temperature ($T_e$), and deprojected galactocentric distance ($R_g/R_{25}$)\footnote{All PAH intensity ratios have been examined and referenced depending on different behavior to the 6.2/11.2 PAH intensity ratio.} 

An anti-correlation is observed between the PAH ratios and the hardness of the radiation field, as probed by [\SIV]/[\NeII] (see Appendix \ref{appsec:NeonSulfurCorr}), with the 6.2/11.2 having the tightest relation. An anti-correlation is also observed between the PAH intensity ratios with oxygen abundance ($12 + \mathrm{log(O/H)}$), a tracer for metallicity, with the 6.2/11.2 ratio showing the strongest relation. Finally, a systematic increase of the PAH ratios with $T_{e}$ in the ionized gas (calculated based on the [N\,\textsc{ii}] $\lambda5575$ emission line; C15) is reported. No straightforward dependence of the PAH intensity ratios on $n_{e}$ of the ionized gas (measured by the [S\,\textsc{ii}] $\lambda\lambda6717,6731$ ratio; C15; not shown) was found. 

PAH emission in astrophysical sources is mainly controlled by the PAH abundance, excitation, the degree of ionization (which depends on $n_{e}$, the intensity of the FUV radiation field, and the gas temperature), the size distribution, and the structure of the PAH molecules present. The PAH abundance can be modified through a various formation scenarios and photo-destruction. The ``top-down" formation scenario is controlled by photoprocessing of VSGs \citep[e.g.][]{Berne2007} and dust grains \citep{Plante2002}--which is linked to the intensity and hardness of the radiation field, supernovae shocks \citep{OHalloran2006}, or the collisional fragmentation of carbonaceous grains \citep{Jones1996}. ``A bottom-up" formation scenario in the ISM is currently being explored following the recent detections of small PAHs in the molecular cloud TMC~1 \citep{McGuire2018,McGuire2021,McCarthy2021, Cernicharo:2021}. Furthermore, metallicity can have a twofold impact on PAH abundances. In low-metallicity galaxies, PAHs may be formed less efficiently around evolved stars \cite[e.g.][]{Galliano2008}, but can also be more efficiently subjected to destruction or processing due to exposure to harder and more intense radiation fields \citep{Madden2006, Engelbracht2008, Gordon2008}. 

The observed decrease in the PAH intensity ratios with increasing hardness of the radiation field (Fig. \ref{fig:M51PAHPhysicalProperties}; leftmost panel), when examining ENRs as a group, can be indicative of overall processing or destruction of PAHs, with medium-sized PAHs (i.e., the carriers dominating the 6.2, 7.7, or 8.6 $\mu$m emission) being more affected in harder radiation fields than larger PAHs responsible for the 11.2 $\mu$m emission. Indeed, the observed range in PAH intensity ratios is a factor of $\sim 2$. \cite{Maragkoudakis2018b} described fairly constant 7.7/11.2 ratios with increasing [\NeIII]/[\NeII] ratios for the mild radiation field hardness ([\NeIII]/[\NeII] $<0.2$) of the M83 \HII{} regions, however 6 M33 \HII{} regions showed a factor of $\sim 3$ decrease in their 7.7/11.2 ratio, having comparable radiation fields as those of Seyfert galaxies ([\NeIII]/[\NeII] $>0.2$).

The PAH intensity ratios of the integrated ENR emission show an anti-correlation with oxygen abundance, especially for the 6.2/11.2 and 8.6/11.2 ratios (Fig. \ref{fig:M51PAHPhysicalProperties}; second panel from the left). Assuming that the hardness of the radiation field increases in lower metallicity environments, the observed decrease of the PAH intensity ratios with increasing oxygen abundance implies variations in the formation of PAHs, with larger PAHs forming more efficiently in regions of higher metallicity. \cite{Maragkoudakis2018b} found no significant variations between the PAH intensity ratios in M33 with respect to elemental abundances as measured by the sulphur to hydrogen (S/H) ratio. \cite{Gordon2008} who studied trends in the PAH behavior as a function of ionization hardness and metallicity in the \HII{} regions of M101, concluded that the decrease in PAH equivalent widths, at least in massive star-forming regions, is primarily due to processing of the dust grains as a result of the radiation field hardness, and to a second extend due to formation, i.e., metallicity, given the better correlation of PAH equivalent widths with the ionization index (a measure of radiation hardness). Although both processing (due to radiation field hardness) and formation (through metallicity) regulate PAH emission, the higher variance in the measured PAH intensity ratios of M51a's ENRs with increasing radiation field hardness indicates that the radiation field hardness has a higher impact on PAH emission, in agreement with \cite{Gordon2008}.

\subsubsection{Dependence with galactocentric distance} 
\label{sec:Rg-dependence}
A positive correlation between all PAH intensity ratios with normalized galactocentric radius ($R_g/R_{25}$) is observed for both the integrated ENR and SF regions (Fig. \ref{fig:M51PAHPhysicalProperties}; rightmost panel, for the 6.2/11.2 case). The increase of the PAH intensity ratios with galactocentric distance $R_{g}$ can be examined as a function of the galaxy's metallicity gradient \citep{Moustakas2010, Croxall2015} as well as the variation of the hardness of the radiation field with $R_{g}$. The oxygen abundance ($12 + \mathrm{log(O/H)}$) gradient yields that the metallicity decreases at higher $R_{g}$. Examining the hardness of the radiation field with $R_{g}/R_{25}$  we only see a subtle decrease of [S\,\textsc{iv}]/[Ne\,\textsc{ii}] albeit with significant scatter, for both the integrated ENRs or their SF component. \cite{Maragkoudakis2018b} in contrast found a moderate increasing gradient of [\NeIII]/[\NeII] with distance from the center for the \HII{} regions in M33, while they observed no gradient in the \HII{} regions of M83 which has a significantly lower metallicity gradient than M51a and M33. Therefore, in the case of M51a, although the hardness of the radiation field has potentially less impact on PAH processing at larger $R_{g}$, the increase in the PAH intensity ratios with $R_{g}$ is indicative that metallicity, and subsequently PAH abundance regulates this increase, and specifically variations in the size distribution of PAHs with a higher ratio of small-to-large PAHs present with increasing distance from the galactic center.


\begin{figure*}
\includegraphics[width=\textwidth]{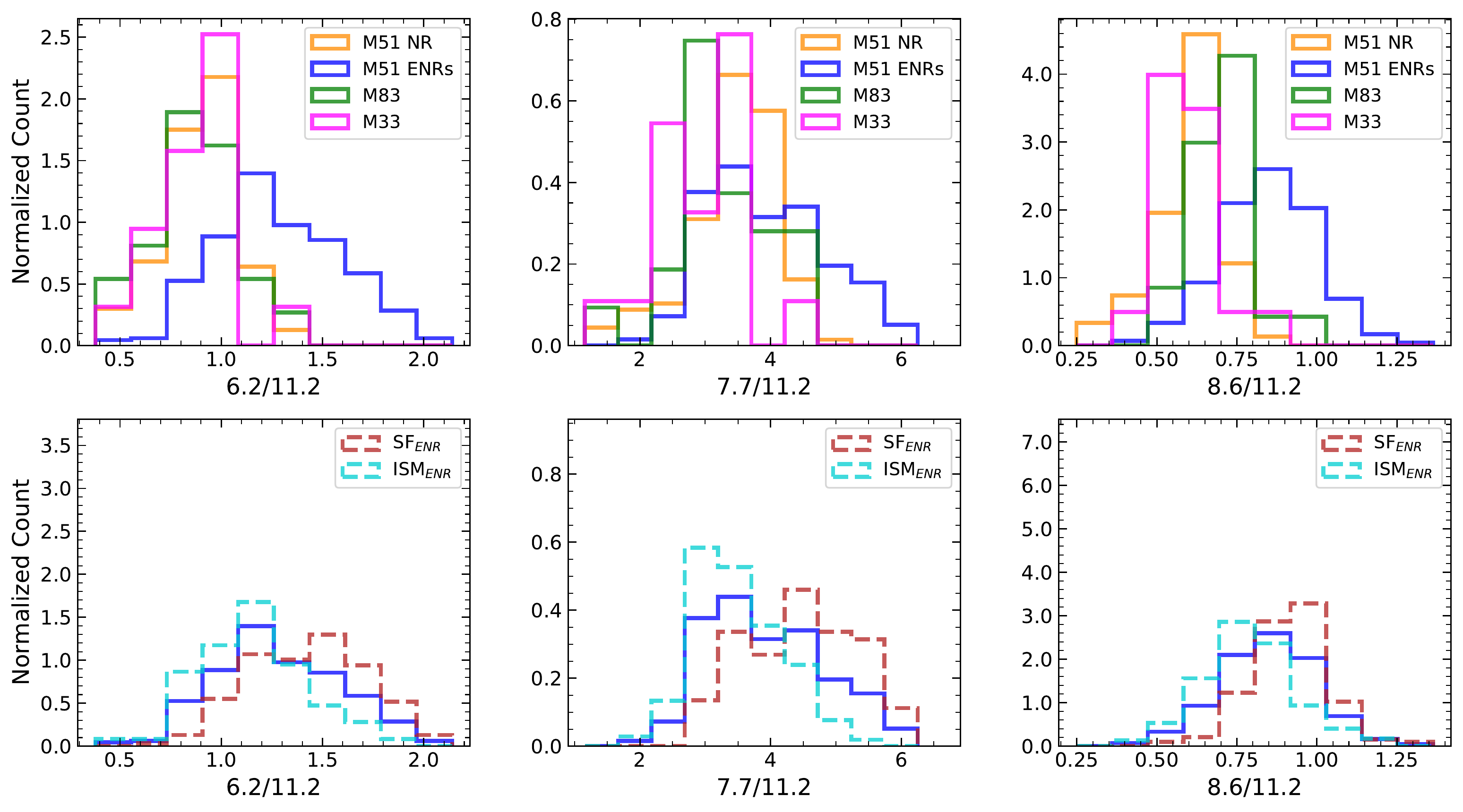}
\caption{The distributions of the 6.2/11.2, 7.7/11.2, and 8.6/11.2 PAH intensity ratios for the M51a NR and ENRs, and the \HII{} regions in M33 and M83 (top) and for the SF and ISM components of M51a ENRs along with the ENRs for comparison (bottom) using 10 bins for each galactic sample and normalized so that area under each distribution equals 1 (i.e. the counts in each bin are divided by the total number of counts in a given sample and multiplied by the bin width). 
}
\label{fig:ratio-histograms112}
\end{figure*}

\subsubsection{Comparison with the M83 and M33 \HII{} regions}
\label{subsec:compM83andM33}

We compared the PAH intensity ratios between the spatially resolved M51a regions in this work and the H\,\textsc{ii} regions in the star-forming galaxies (SFGs) M83 and M33 as examined in \cite{Maragkoudakis2018b}. Although M51a hosts an AGN and therefore the integrated emission features compared to SFGs or resolved \HII{} regions may differ due to intrinsically different environments \citep[e.g.][]{Smith07b, Maragkoudakis2018b}, examination of the PAH emission characteristics between the \HII{} regions in M83 and M33 and M51a's ENRs is analogous. In addition, a comparison of the PAH emission in star-forming regions and the central regions of AGN hosts, such as M51a's NR, can offer insights on the impact of AGN radiation on PAH emission. 

\begin{table*}
    \begin{center}
    \caption {K-S test $p-$values of the comparison between the intensity distributions of M51a NR, M51a ENRs, M33 \HII{} regions, and M83 \HII{} regions.} 
    \label{tab:M51KSTest}
    \begin{tabular}{@{}lcccccccc}
    
    \hline
    Region Comparison&6.2/11.2&7.7/11.2&8.6/11.2&7.7/6.2&8.6/6.2&7.7/8.6&log(PAH/VSG)\\
    \hline
    NR / ENRs&2.56E-36&2.14E-08&3.20E-57&5.63E-42&6.98E-02&6.66E-16&1.38E-14\\ 
    NR / M83&0.27&0.14&1.76E-04&0.84&2.20E-06&1.16E-10&3.33E-16\\ 
    NR / M33&0.73&1.27E-03&0.31&4.47E-10&0.90&1.71E-05&2.22E-16\\ 
    ENRs / M83&1.15E-07&6.13E-03&1.56E-06&2.85E-12&3.03E-06&0.32&2.22E-16\\ 
    ENRs / M33&9.50E-09&3.40E-06&5.33E-12&0.05&0.22&0.48&1.80E-13\\ 
    M83 / M33&0.91&0.13&2.70E-03&1.47E-07&5.75E-03&0.12&2.14E-05\\
    \hline
\end{tabular}
\end{center}
\end{table*}

Examination of the distribution of the ionization fraction as described independently by the distributions of the 6.2/11.2, 7.7/11.2, and 8.6/11.2 intensity ratios (Fig. \ref{fig:ratio-histograms112}) reveals similarities and differences across the different type of sources, depending on the intensity ratio used as a proxy of the ionization fraction. In all three PAH intensity ratio distributions, M51a ENRs have distinct distributions exceeding to higher ionization fractions compared to M83 and M33 \HII{} regions and the NR. A breakdown of ENRs into their SF and ISM components (Fig. \ref{fig:ratio-histograms112}, bottom row) reveals that although both the SF and ISM ENRs span similar PAH intensity ratio ranges the tail of the ENRs distribution towards higher values has a higher contribution from the SF regions. The 6.2/11.2 intensity ratio distributions indicate that M51a NR has similar ionization fractions to M83 and M33 \HII{} regions ($p-$values of 0.27 and 0.73 as reported by a K-S test; see Table \ref{tab:M51KSTest}). On the other hand, the 7.7/11.2 distributions show that M51a NR and M83's \HII{} regions have similar ionization fractions ($p-$value = 0.14), while the 8.6/11.2 ratio suggests that M51a NR and M33's \HII{} regions have similar distributions ($p-$value = 0.31). 

\begin{figure*}
\includegraphics[width=\textwidth]{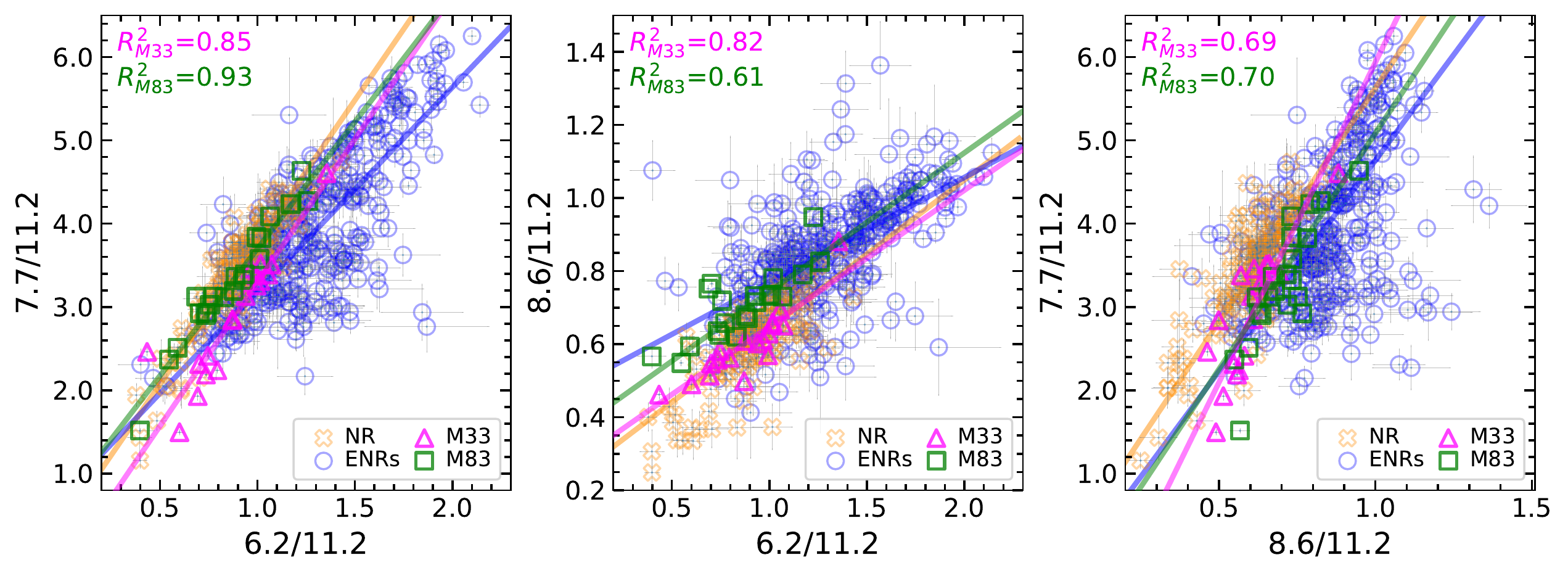}
\includegraphics[width=\textwidth]{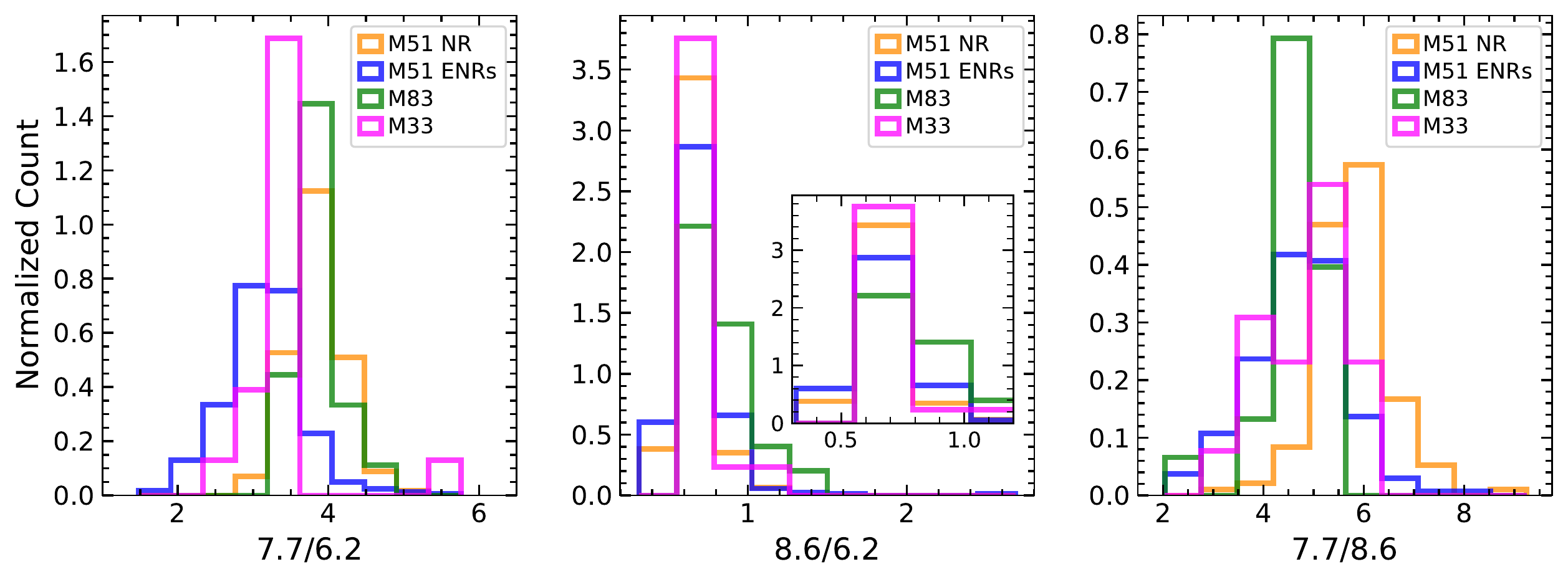}
\caption{Top row: Comparison among the 7.7/11.2, 6.2/11.2, and 8.6/11.2 PAH intensity ratios of M51a's NR (orange crosses) and ENRs (blue circles), and the star-forming regions of M83 (green squares) and M33 (pink triangles). Linear regression fits and regression coefficients ($R^2$) are displayed for each case in respective colors (see Fig.~\ref{fig:M51-Overview} for $R^2$ values of M51a). Bottom row: The distributions of the 7.7/6.2, 8.6/6.2, and 7.7/8.6 ratios (slopes of the top-row plots). Normalization was done in the same procedure as described in Fig. \ref{fig:ratio-histograms112}.}
\label{fig:M51M33M83}
\end{figure*}

\begin{figure}
\includegraphics[width=0.45\textwidth]{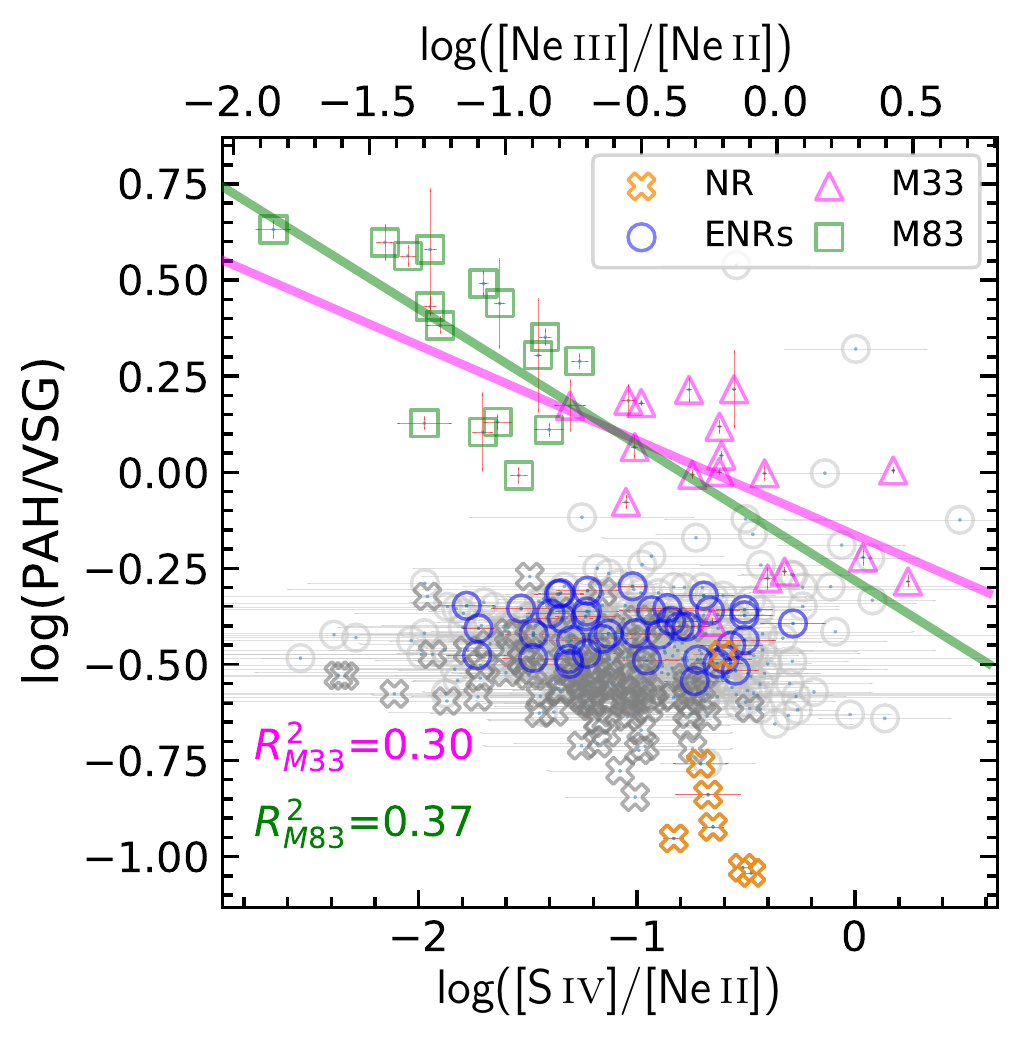}
\caption{The PAH/VSG as a function of [S\,\textsc{iv}/Ne\,\textsc{ii}] for the M51a NR and ENRs, and the \HII{} regions of M33 and M83. Linear regression fit and regression coefficients ($R^2$) are presented for the M33 and M83 \HII{} regions. Errors from the above regions are represented by red lines. Gray-scale points and error lines represent pixels from M51a NR or ENR without a 3-sigma detection in the [S\,\textsc{iv}]/[Ne\,\textsc{ii}] ratio.
The log([N\,\textsc{iii}]/[Ne\,\textsc{ii}]) values in the top axis for all data points are estimated as described in Appendix \ref{appsec:NeonSulfurCorr}. }
\label{fig:M51VSGSIV}
\end{figure}

\begin{figure*}
\includegraphics[width=0.9\textwidth]{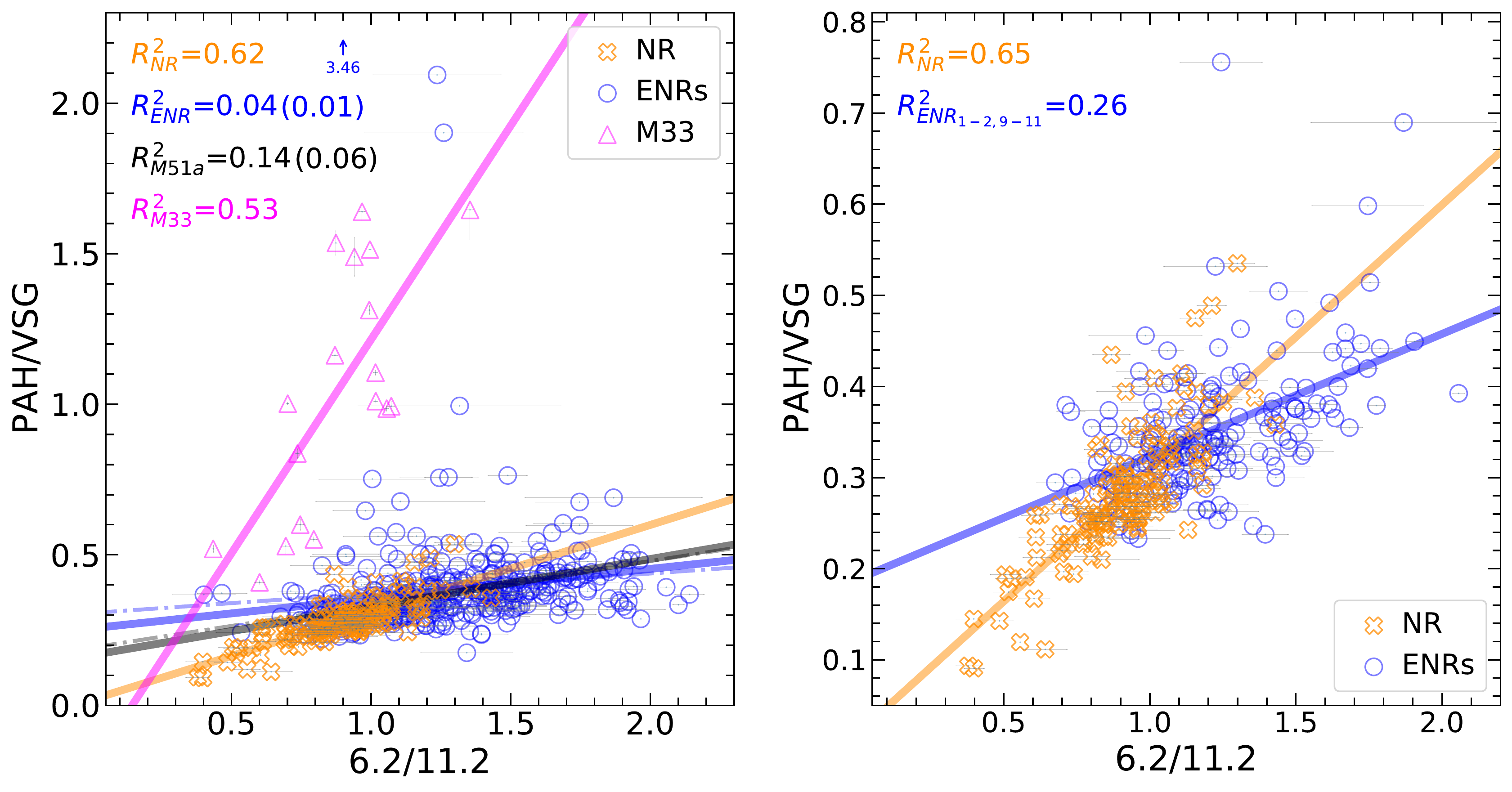}
\caption{PAH/VSG as a function of 6.2/11.2. Left: The relation for M51a NR, ENRs, and M33. The $R^2$ value and line of best-fit for M51a NR are shown in orange while the $R^2$ value and line of best-fit for M51a ENRs, excluding the outlier indicated by blue arrow, are shown in blue. The best-fit line and $R^2$ value for the entire M51a is shown in black. Dash-dotted lines and $R^2$ values in brackets represent analysis of M51a and ENRs including the outlier point indicated by a blue arrow. The best-fit line and $R^2$ value for M33 are given in magenta. Right: the relation for M51a NR and those ENRs showing correlation based on the Pearson correlation test as per Table \ref{tab:M51pearson}.}
\label{fig:M51VSG62}
\end{figure*}

Fig. \ref{fig:M51M33M83} presents the relative PAH intensity correlations for the M51a NR and ENRs, with M33 and M83 \HII{} regions overplotted (top row), and their respective slope distributions (bottom row). In the 7.7/11.2 -- 6.2/11.2 plane (Fig. \ref{fig:M51M33M83}, top left panel) the M83's \HII{} regions coincide with the M51a NR pixels, having similar slopes (7.7/6.2) of $3.05 \pm 0.20$ and $3.41 \pm 0.11$ respectively (Table \ref{Apptab:M51slopePAHRatios}), while M33's regions are offset towards lower 7.7/11.2 values with a slope (7.7/6.2) of $3.44 \pm 0.30$. This is further demonstrated when examining the respective distributions of the 7.7/6.2 ratio where M83's regions and M51a NR have similar distributions (K-S test $p-$value = 0.84). In the 8.6/11.2 -- 6.2/11.2 plane (Fig. \ref{fig:M51M33M83}, top middle panel) the M33 \HII{} regions are now those overlapping with the M51a NR pixels, with respective slopes of $0.37 \pm 0.04$ and $0.41 \pm 0.03$ and similar distributions, while M83's regions have similar (8.6/6.2) slope to M51a's ENRs being $0.38 \pm 0.07$ and $0.29 \pm 0.01$ respectively. In addition, M51a NR and M33 \HII{} regions have similar 8.6/6.2 distributions (K-S test $p-$value = 0.90). In the 7.7/11.2 -- 8.6/11.2 plane (Fig. \ref{fig:M51M33M83}, top right panel) the M83 and M33 \HII{} regions coincide with the bulk of M51a ENRs, having similar 7.7/8.6 distributions (K-S test $p$-value of 0.32 and 0.48 respectively).

The similarity between the 6.2/11.2 distributions in M51a NR, M83 and M33 \HII{} regions is suggestive of a similar 6.2 \micron{} emission distribution--as normalized based on their 11.2 \micron{} emission--respectively among these regions. At the same time, the similarity of the M51a NR and M83's \HII{} region 7.7/6.2 distribution, and the M51a NR and M33's \HII{} region 8.6/6.2 distribution is indicative that the M51a NR has similar 7.7 \micron{} PAH emission carriers to M83, and similar 8.6 \micron{} PAH emission carriers to M33, considering that all three region groups have comparable 6.2 \micron{} distributions. 

In general, the 8.6 \micron{} emission is attributed to large (100--150 carbon atoms), compact, symmetric PAHs \citep{Bauschlicher_Jr__2008, Bauschlicher_2009, Peeters2017}. Regarding the 7.7 \micron{} complex emission, this can be decomposed into two Gaussians components, G7.6 and G7.8 centered at approximately 7.6 and 7.8 \micron{} respectively, with different carrier characteristics \citep[e.g.][]{Peeters2017}. Specifically, the G7.6 \micron{} emission is primarily assigned to compact, ionized PAHs with 50-100 carbon atoms, while the G7.8 \micron{} emission is attributed predominantly to neutral, irregular, large PAHs (100--150 carbon atoms), or PAH clusters with bay regions. As noted by \cite{Bauschlicher_2009}, while large irregular PAHs exhibit features at 7.8 \micron{}, the 8.6 \micron{} band in astrophysical sources arises exclusively from large, compact PAH cations and anions.

The similarities among the 7.7 and 8.6 \micron{} PAH emission features in M51a NR with M83 and M33 respectively, can thus be explained considering the respective carrier characteristics (i.e., size and symmetry) as well as the radiation field properties in the respective sources. The aromatic emission in M51a NR is the product of a mixture of PAHs of different sizes and symmetries as shaped by the radiation field of the central source, considering that the $60'' \times 33''$ NR aperture encompasses emission from regions at the vicinity of the AGN but also from regions further out at the circumnuclear area of the galaxy. Consequently, the similarity between the 8.6 \micron{} emission in M51a NR and M33 can be attributed to pixels/regions populated mostly by large compact PAHs, while the similarity in the the 7.7 \micron{} emission between M51a NR and M83 can be due to pixels dominated by large irregular PAH populations. Furthermore, M33 \HII{} regions have harder radiation fields, as parametrized by the [Ne\,\textsc{iii}]/[Ne\,\textsc{ii}] ratio, compared to those of M83, covering the same [Ne\,\textsc{iii}]/[Ne\,\textsc{ii}] range as AGN galaxies \citep{Maragkoudakis2018b}. Pixels lying close to the vicinity of the AGN are expected to have comparable hard radiation fields as those in M33's \HII{} regions, and for those regions the smaller G7.6 carriers would be more subjective to destruction compared to the G7.8 and G8.6 carriers, resulting in the observed similarity between the M51a NR and M33 8.6 (or 8.6/6.2) \micron{} distributions as well as in the difference between their respective 7.7 (or 7.7/6.2) \micron{} distributions. In addition, compact PAHs--such as the carriers of the 8.6 \micron{} emission-- are generally more stable and can sustain radiation field strength compared to irregular PAH carriers. Likewise, M51a NR pixels further away from the AGN will be subjected to less hard radiation fields as those of M83's \HII{} regions, where both the different carriers of the 7.7 \micron{} complex emission will be present, including irregular PAHs. 

Summarizing, our results indicate similarities between: (i) the 7.7 \micron{} carriers in M51a's NR and M83's \HII{} regions; (ii) the 8.6 \micron{} carriers in M51a's NR and M33 \HII{} regions; (iii) both type of carriers between M51a's ENRs and M33's and M83's \HII{} regions.

\subsection{PAH/VSG}
\label{subsec:PAH_VSGData}

Continuum emission longward of 10 \micron{} is generally attributed to very small grains \citep[VSGs;][]{Desert1990}. Observations of resolved SF regions showed that VSG emission peaks within \HII{} regions, whereas PAH emission dominates in photo-dissocation regions \citep[PDRs, e.g.][]{Bendo2008, Gordon2008}. Consequently, low PAH/VSG ratios are typically measured in \HII{} regions. VSG emission is also detected in the ISM however at higher PAH/VSG ratios than in \HII{} regions \citep[e.g.][]{Bendo2008}. These variations in the PAH/VSG ratio depend on the abundance of the PAH and VSG carriers, but are also directly linked to the characteristics of the radiation field. Specifically, 
the PAH/VSG ratio anti-correlates with the hardness of the radiation field as traced by [\NeIII]/[\NeII] \citep[e.g.][]{Madden2006,Lebouteiller_2007, Maragkoudakis2018b}. In this section, we discuss the relation of the PAH/VSG ratio with the hardness of the radiation field (Section~\ref{sec:PAHVSG_hardI}) and with the PAH intensity ratios (Section~\ref{sec:PAHVSG_PAHratios}).

\subsubsection{PAH/VSG and the radiation field hardness}
\label{sec:PAHVSG_hardI}
We examined the variation of the PAH/VSG ratio in the M51a NR and ENRs as a function of the hardness of radiation field, described by the [S\,\textsc{iv}]/[Ne\,\textsc{ii}] ratio (see Appendix \ref{appsec:NeonSulfurCorr}), further including the \HII{} regions of M83 and M33 for comparison. The VSG emission was defined as the integrated dust continuum emission between 10 and 14 \micron{}, and the total PAH emission as the sum of the 6.2, 7.7, 8.6, 11.2, and 12.6 \micron{} feature intensities. The previously reported decrease of the PAH/VSG ratio in M83 and M33 \HII{} regions with increasing [Ne\,\textsc{iii}]/[Ne\,\textsc{ii}] \citep{Maragkoudakis2018b}, is also evident when adopting the [S\,\textsc{iv}]/[Ne\,\textsc{ii}] ratio as tracer of the radiation field hardness (Fig. \ref{fig:M51VSGSIV}). 
However, we do not observe a similar anti-correlation for M51a ENRs while M51a NR displays unique behavior. Specifically, the M51a NR has the lowest PAH/VSG values which gradually decrease with increasing [S\,\textsc{iv}]/[Ne\,\textsc{ii}], designating hard, narrow-ranged local radiation fields ($-0.83<$ log([S\,\textsc{iv}]/[Ne\,\textsc{ii}]) $<-0.48$), potentially associated with the continuum emission from the AGN. In contrast, the PAH/VSG ratio of M51a ENRs shows a similar range of variation to that of the NR ($\Delta\log(\mathrm{PAH/VSG}) \sim 0.25$) but throughout a much wider range in [S\,\textsc{iv}]/[Ne\,\textsc{ii}] values ($-1.85<$ log([S\,\textsc{iv}]/[Ne\,\textsc{ii}]) $<-0.4$), even when considering only the SF pixels (see Section \ref{subsec:sfvsism}). 
ISOCAM observations of M51a presented in \cite{Madden2006} display, on average, a slight decrease in PAH/VSG with increasing radiation field hardness albeit with significant scatter. While the data presented here cover roughly the same range and distribution in radiation field hardness as the ISOCAM data, our PAH/VSG ratios only sample the lower end of the PAH/VSG range seen in the ISOCAM data (after correction for the different adopted ranges for the VSG measurements\footnote{As the IRS-SL observations cover the 5--15$\mu$m, we adopted a 10 to 13.5 \micron{} range for the VSG measurement where as \citet{Madden2006} used a 10-16 $\mu$m range, which affects the measured intensity. Based on ISO-SWS observations of the Orion Bar, \citet{Knight:orion} estimated a difference of a factor $\sim 2.2$ for the VSG intensity calculation due to this different wavelength range. In addition, we use a different spectral decomposition method to extract the PAH emission, further complicating a direct comparison with \citet{Madden2006}.}). 
We note that all resolved sources presented in \citet{Madden2006} show significant scatter in the PAH/VSG vs [Ne\,\textsc{iii}]/[Ne\,\textsc{ii}] relationship with many not exhibiting a clear anti-correlation, however when compared with their entire sample, an overall anti-correlation becomes apparent \citep{Madden2006}. 

The lack of a clear anti--correlation between the PAH/VSG ratio and the hardness of the radiation field within the M51a data
is consistent with the results from \citet[][their Figure 16]{Madden2006} where the strongest anti--correlation within an extended source (by far) was observed for the nearby Galactic \HII{} region M17. However, although a fraction of M33 \HII{} regions and ENRs cover similar ranges in [S\,\textsc{iv}]/[Ne\,\textsc{ii}] as the NR, signifying similar radiation fields, their respective PAH/VSG values are 
significantly different at a given [S\,\textsc{iv}]/[Ne\,\textsc{ii}] value. The same holds for M51a ENRs and a fraction of the M33 and M83 \HII{} regions. This suggests a difference in the relative abundance of PAHs and VSG among the different regions and sources, that is thus not driven by the hardness of the radiation field.

\subsubsection{The connection of PAH/VSG with PAH intensity ratios}
\label{sec:PAHVSG_PAHratios}
Our analysis further identified a positive linear correlation between PAH/VSG and the 6.2/11.2, 7.7/11.2, and 8.6/11.2 PAH intensity ratios respectively, as mapped by the M51a NR and ENRs, and M33 \HII{} regions (Fig. \ref{fig:M51VSG62}, left panel; weighted Pearson's coefficients are given in Table \ref{tab:M51pearson}). No correlation was observed in M83 \HII{} regions. The M33 \HII{} regions have the steepest slope $\alpha = 1.42 \pm 0.34$ (Table \ref{Apptab:M51slopePAHVSG}) compared to the M51a NR and ENRs. M51a NR has a slope of $0.29 \pm 0.02$ while PAH/VSG for ENRs are fairly constant with increasing 6.2/11.2 with a slope of $0.10 \pm 0.02$. M51a as a whole--considering the NR and ENR pixels combined--has a weighted Person's coefficient of $0.39$ and a slope of $0.16 \pm 0.02$. A separate examination of the individual M51a ENRs reveals that such a correlation is present in certain ENRs while absent in others (Table \ref{tab:M51pearson}). The right panel of Fig. \ref{fig:M51VSG62} shows the PAH/VSG -- 6.2/11.2 correlation for ENRs with a weighted Pearson's coefficient above 0.4 exhibiting now a high positive correlation ($p_{w}=0.57$) similar to M51a's NR ($p_{w}=0.78$). Similar behaviour of the the PAH/VSG correlation with the ionization fraction is observed when using other tracers such as 7.7/11.2 and 8.6/11.2 PAH intensity ratios (see Appendix \ref{appsec:VSGPAHRatioCorr}).
Finally, we do not observe a correlation between the PAH/VSG ratio and the metallicity.

The increase of the PAH/VSG ratio with increasing ionization fraction is, to a first degree, related to the increase of the total PAH emission which consist of--based on its current definition in this work--the sum of five PAH emission features (at 6.2, 7.7, 8.6, 11.2, and 12.7 \micron), three of which are attributed principally to ionized PAHs (6.2, 7.7, and 8.6 \micron{} features), one feature attributed to both neutral and ionic PAHs (12.7 \micron{} feature), and one feature attributed to neutral PAHs (11.2 \micron{} feature). The different slopes observed for M51a and M33, as well as the fact that no correlation is seen for the M83 \HII\, regions, confirms that the PAH ionization fraction is not the main driver of the variations observed in the PAH/VSG ratio.
For a better probe of the PAH/VSG dependence with the ionization fraction of PAHs, a larger sample of diverse sources would be required.

\section{Summary and conclusions}
\label{sec:Summary_conclusion}

We performed a detailed examination of the spatially resolved mid-IR characteristics in the M51a (circum)nuclear region and 11 extranuclear regions using \textit{Spitzer}-IRS observations, and performed a comparison between the M51a PAH emission features and VSG emission, with those in the extranuclear regions of the star-forming galaxies M33 and M83. The main conclusions of this work are summarized as follows.

(i) The M51a NR and ENRs exhibit correlations among the PAH intensity ratios which describe the ionization fraction, with the NR having the firmest correlation, while ENRs as a whole showing increased scatter.

(ii) Separate examination of the ENRs after separating into star-forming (SF) and ISM dominated pixels, revealed that it is the ISM pixels that introduce the observed scatter in the correlation plots of all ENRs, while SF pixels have firm correlations.

(iii) The dependence of the 6.2/11.2, 7.7/11.2, and 8.6/11.2 PAH intensity ratios in M51a's ENRs to their local physical conditions and galactocentric radii revealed the following: There is an anti-correlation of the PAH intensity ratios with oxygen abundance ($12 + \mathrm{log(O/H)}$) as well as the hardness of the radiation field, as described by [\SIV]/[\NeII], and a positive correlation with $R_{g}/R_{25}$. The decrease of the PAH intensity ratios with the hardness of the radiation field is suggestive of overall processing of PAHs, however the dependence with $12 + \mathrm{log(O/H)}$ suggests a higher rate of small PAH processing compared larger PAHs. The increase of PAH intensity ratios with $R_{g}/R_{25}$ are indicative of variations in the size distribution of PAHs, with a higher ratio of small-to-large PAHs formed with increasing distance from the galactic center. 

(iv) M51a's NR has similar 7.7/6.2 slope with the M83 and M33 \HII{} regions, and similar 8.6/6.2 distributions with M33's \HII{} regions, while M51a ENRs have comparable 7.7/8.6 distributions with both M83 and M33 \HII{} regions. Given the analogous 6.2 \micron{} distribution among the different sources as normalized to the 11.2 \micron{} feature, the similarities among the previous distributions are indicative of similarities between the 7.7 \micron{} carriers in M51a's NR and M83's \HII{} regions, the 8.6 \micron{} carriers in M51a's NR and M33 \HII{} regions, and both type of carriers between M51a's ENRs and M33's and M83's \HII{} regions.

(v) The ratio of PAH/VSG in M51a shows no clear anti-correlation, with only the NR regions decreasing at high [\SIV]/[\NeII] values ($> -1$), while the ENRs exhibit a nearly constant PAH/VSG ratio through a wide range of [\SIV]/[\NeII] values. For a given [\SIV]/[\NeII] ratio, significantly different PAH/VSG ratios are observed indicating a difference in relative abundance of PAHs and VSG that is not driven by the hardness of the radiation field.

(vi) We have identified a positive linear correlation between PAH/VSG and the 6.2/11.2, 7.7/11.2, and 8.6/11.2 PAH intensity ratios in M51a and M33 \HII{} regions, indicative of a correlation of the PAH/VSG ratio with the PAH ionization fraction. The lack of such correlation in M83 or certain M51a ENRs, indicates that the PAH ionization is not the main driver of the PAH/VSG ratio.

\section*{Acknowledgments}
    We would like to thank the referee for the constructive comments and suggestions that have improved the clarity of this paper. R.X.Z.'s research was partially supported by an Ontario Graduate Scholarship. A.M.'s research was partially supported by an appointment to the NASA Postdoctoral Program at NASA Ames Research Center, administered by the Universities Space Research Association under contract with NASA. E.P. acknowledges support from an NSERC Discovery Grant (RGPIN-2020-06434). 

\section*{Data Availability Statement}
    The analysis products of this work will be shared on a reasonable request to the corresponding author.

\bibliographystyle{mnras}
\bibliography{Bibliography}

\appendix

\section{Stitching Factor and influence on regression analysis}
\label{appsec:StichingFacorRegression}

\begin{figure*}
    \includegraphics[width=\textwidth]{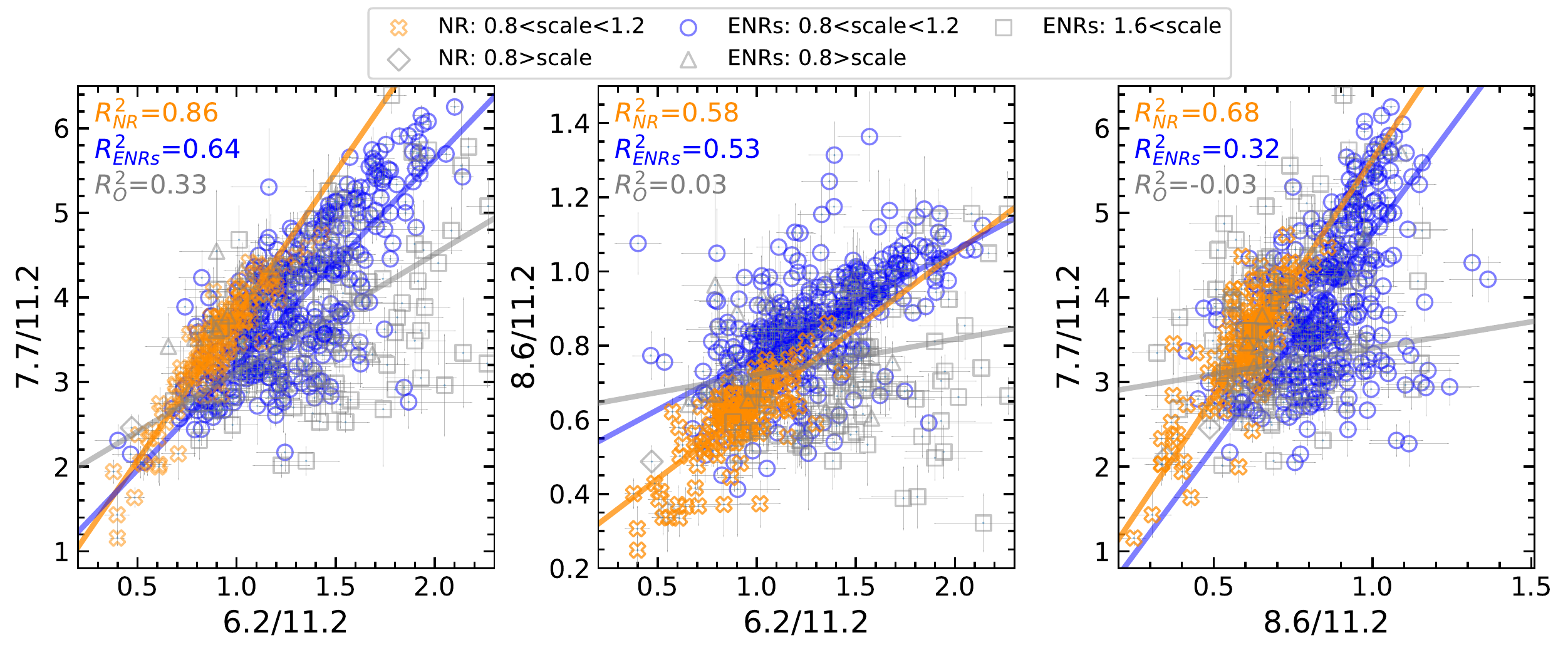}
    \caption{Relative comparison among the 7.7/11.2. 6.2/11.2, and 8.6/11.2 PAH intensity ratios of the M51a NR and ENRs (points and lines as described in Fig. \ref{fig:M51-Overview}), further including rejected pixels (shown in gray) outside the adopted stitching factor range (see Section \ref{subsec:dataReduction}).}
    \label{appfig:intensitystitch}
\end{figure*}

A fraction of M51a ENR pixels with stitching factors outside the adopted 0.8--1.2 range was excluded from the analysis (Section \ref{subsec:dataReduction}). Fig. \ref{appfig:intensitystitch} presents the regression analysis including now the discarded pixels. Inclusion of pixels requiring stitching factor $>1.2$ resulted in a noticeable increase of the 6.2/11.2 intensity ratio due to the artificial enhancement of the 6.2 \micron{} PAH feature. In all PAH intensity ratio correlation plots, the inclusion of pixels outside the adopted stitching factor boundaries caused a decrease in the slope of the linear regression fit. 

\section{Radiation field hardness proxies}
\label{appsec:NeonSulfurCorr}

\begin{figure}
    \includegraphics[width=0.5\textwidth]{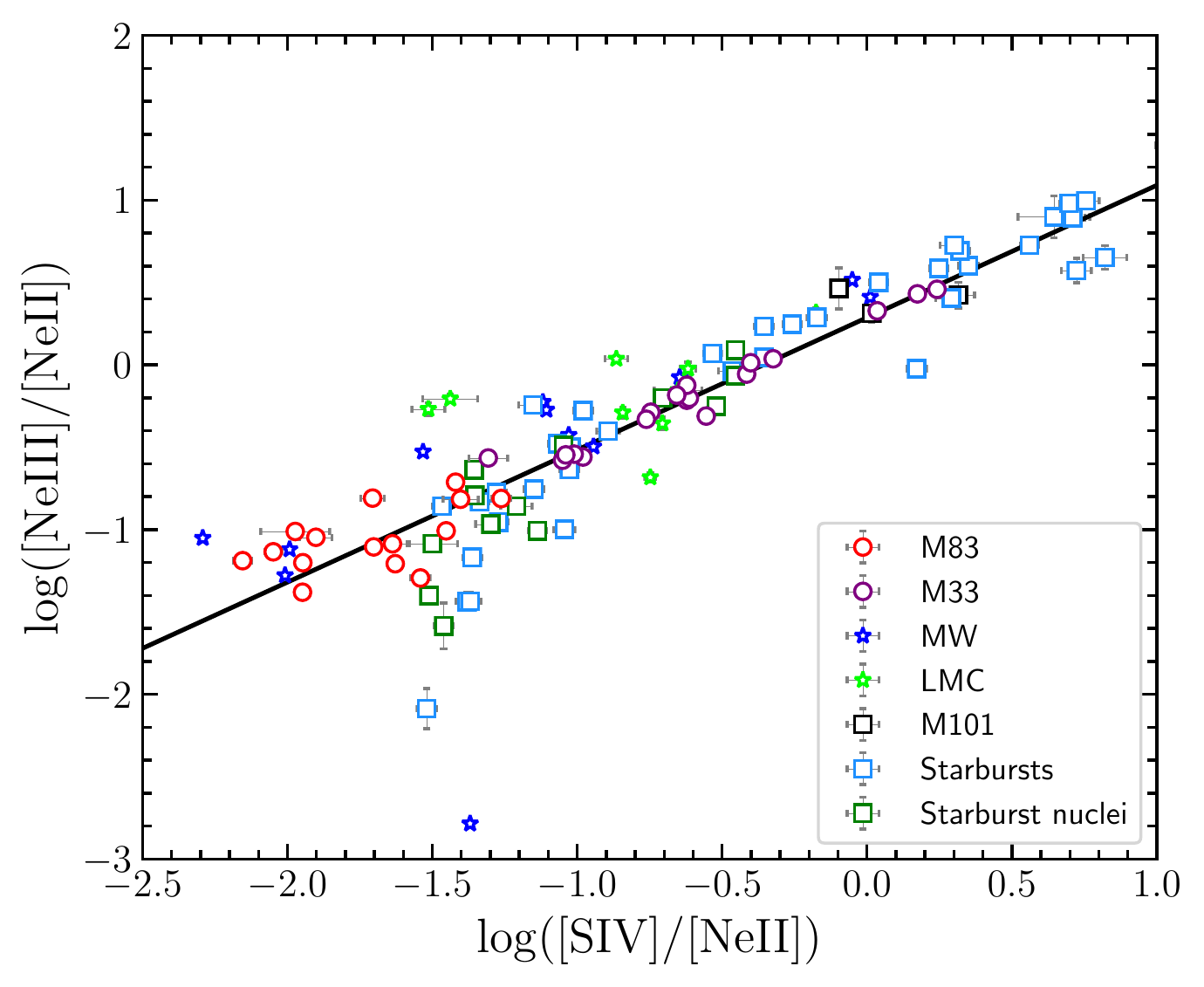}
    \caption{The relation between [NeIII]/[Ne\,\textsc{ii}] and [S\,\textsc{iv}]/[Ne\,\textsc{ii}] for different galactic and extragalactic sources, as described in \protect\cite{Maragkoudakis2018b}. The black line is the linear regression fit to the data.}
    \label{appfig:galaxyNeIII}
\end{figure}

Intensity ratios of [Ne\,\textsc{iii}]/[Ne\,\textsc{ii}] or [S\,\textsc{iv}]/[S\,\textsc{iii}] are the most commonly used tracers of the radiation field hardness. In studies where the [Ne\,\textsc{iii}] and [S\,\textsc{iii}] (at 18.7 or 33.5 \micron) are not covered, the [S\,\textsc{iv}]/[Ne\,\textsc{ii}] ratio is employed as an alternate tracer \citep[e.g.,][]{Lebouteiller_2007}. In Fig. \ref{appfig:galaxyNeIII}, we further demonstrate the strong correlation between these two tracers for the sample of \cite{Maragkoudakis2018b}. We use the linear regression fit to estimate the [Ne\,\textsc{iii}]/[Ne\,\textsc{ii}] values provided in the top axis of Fig. \ref{fig:M51VSGSIV}. 

\section{PAH/VSG correlation with 7.7/11.2 and 8.6/11.2 PAH feature ratios}
\label{appsec:VSGPAHRatioCorr}
\begin{figure*}
\includegraphics[width=0.9\textwidth]{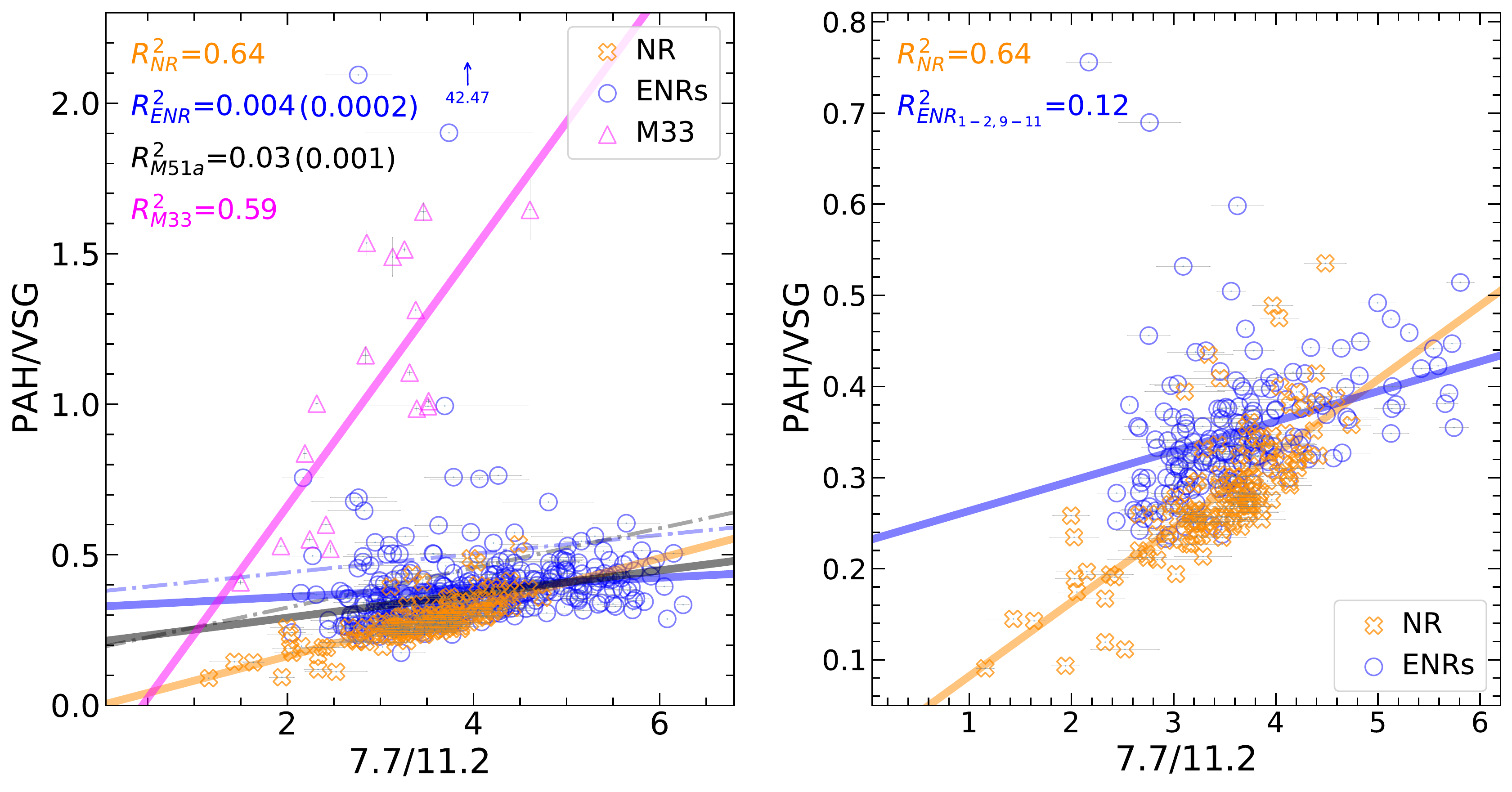}
\includegraphics[width=0.9\textwidth]{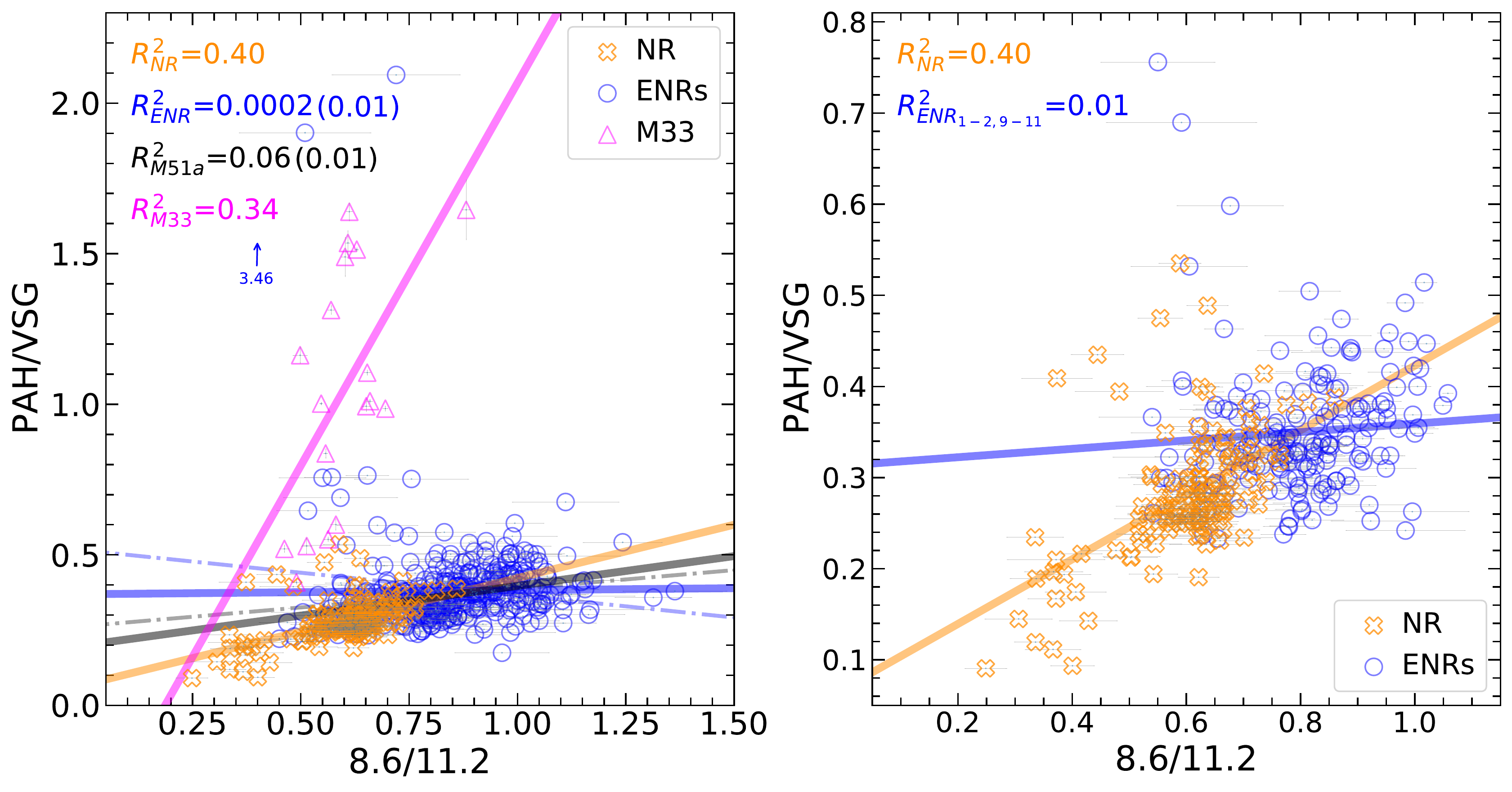}
\caption{PAH/VSG as a function of 7.7/11.2 (top row) and 8.6/11.2 (bottom row). Left panels: The relation for M51a NR, ENRs, and M33. The $R^2$ value and line of best-fit for M51a NR are shown in orange while the $R^2$ value and line of best-fit for M51a ENRs, excluding the outlier indicated by blue arrow, are shown in blue. The best-fit line and $R^2$ value for the entire M51a is shown in black. Dash-dotted lines and $R^2$ values in brackets represent analysis of M51a and ENRs including the outlier point indicated by a blue arrow. The best-fit line and $R^2$ value for M33 arg given in magenta. Right panels: the relation for M51a NR and those ENRs showing correlation based on the Pearson correlation test as per Table \ref{tab:M51pearson}.}
\label{appfig:M51VSG7786}
\end{figure*}

We reported a weak positive correlation between the PAH/VSG ratio and various PAH ratios in section \ref{subsec:PAH_VSGData}. 
Figure~\ref{appfig:M51VSG7786} shows the PAH/VSG correlation with 7.7/11.2 and 8.6/11.2. The correlation coefficients along with linear regression model are provided in Tables~\ref{tab:M51pearson} and \ref{Apptab:M51slopePAHVSG}.

\section{Correlation parameters}

Here we summarize the linear regression coefficients from the various PAH intensity ratio correlation plots (Table \ref{Apptab:M51slopePAHRatios}), the PAH intensity ratios as a function of M51a local physical parameters (Table \ref{apptab:phys_prop_slopes}), and the PAH/VSG correlations among different sources (Table \ref{Apptab:M51slopePAHVSG}). Table \ref{apptab:M51fluxPropertiesRandp} gives the correlation statistics for PAH ratios and physical properties. 

\begin{table*}
\begin{center}
\caption {Slopes ($\alpha$) and intercepts ($\beta$) of linear regression fits for PAH intensity ratios correlations among different regions and sources.} 
\label{Apptab:M51slopePAHRatios}
   
\begin{tabular}{@{}lccccccc}
\hline
Regions &\multicolumn{2}{c}{7.7/11.2 vs 6.2/11.2}&\multicolumn{2}{c}{8.6/11.2 vs 6.2/11.2}&\multicolumn{2}{c}{8.6/11.2 vs 7.7/11.2}\\
        &$\alpha$&$\beta$&$\alpha$&$\beta$&$\alpha$&$\beta$\\
\hline
NR&3.41$\pm$0.11&0.37$\pm$0.10&0.41$\pm$0.03&0.24$\pm$0.03&5.61$\pm$0.29&0.02$\pm$0.18\\
ENRs (all)&2.45$\pm$0.09&0.74$\pm$0.13&0.29$\pm$0.01&0.48$\pm$0.02&5.06$\pm$0.30&-0.30$\pm$0.27\\
ENR 1&3.16$\pm$0.65&-0.34$\pm$0.83&0.36$\pm$0.13&0.31$\pm$0.18&6.18$\pm$1.61&-1.16$\pm$1.26\\
ENR 2&2.58$\pm$0.20&-0.12$\pm$0.31&0.31$\pm$0.04&0.42$\pm$0.06&4.86$\pm$0.59&-0.52$\pm$0.53\\
ENR 3&2.60$\pm$0.19&0.75$\pm$0.34&0.40$\pm$0.04&0.28$\pm$0.07&5.19$\pm$0.60&0.21$\pm$0.59\\
ENR 4&2.38$\pm$0.12&1.43$\pm$0.18&0.21$\pm$0.02&0.59$\pm$0.03&8.23$\pm$0.98&-2.51$\pm$0.89\\
ENR 5&2.36$\pm$0.10&1.45$\pm$0.16&0.27$\pm$0.03&0.54$\pm$0.05&6.30$\pm$0.73&-0.95$\pm$0.70\\
ENR 6&2.76$\pm$0.40&0.74$\pm$0.70&0.20$\pm$0.07&0.63$\pm$0.12&6.11$\pm$2.77&-0.48$\pm$2.75\\
ENR 7&2.19$\pm$0.23&0.79$\pm$0.34&0.02$\pm$0.06&0.97$\pm$0.11&-0.77$\pm$1.39&4.74$\pm$1.45\\
ENR 8&2.70$\pm$0.12&0.40$\pm$0.18&0.27$\pm$0.04&0.52$\pm$0.06&5.22$\pm$0.94&-0.50$\pm$0.87\\
ENR 9&3.04$\pm$0.23&0.23$\pm$0.27&0.36$\pm$0.03&0.40$\pm$0.03&4.59$\pm$0.80&-0.08$\pm$0.65\\
ENR 10&2.95$\pm$0.18&0.29$\pm$0.21&0.32$\pm$0.03&0.46$\pm$0.04&5.15$\pm$0.69&-0.64$\pm$0.57\\
ENR 11&2.58$\pm$0.17&1.23$\pm$0.23&0.40$\pm$0.04&0.29$\pm$0.06&5.34$\pm$0.55&0.27$\pm$0.45\\
ENR SF&2.48$\pm$0.14&0.80$\pm$0.22&0.26$\pm$0.01&0.53$\pm$0.02&5.99$\pm$0.49&-0.97$\pm$0.46\\
ENR ISM&1.56$\pm$0.13&1.62$\pm$0.16&0.25$\pm$0.03&0.51$\pm$0.03&1.52$\pm$0.35&2.22$\pm$0.28\\
M51a&2.26$\pm$0.06&1.07$\pm$0.09&0.37$\pm$0.01&0.34$\pm$0.01&4.18$\pm$0.19&0.55$\pm$0.16\\
M33&3.44$\pm$0.30&-0.15$\pm$0.26&0.37$\pm$0.04&0.28$\pm$0.04&7.68$\pm$1.21&-1.74$\pm$0.72\\
M83&3.05$\pm$0.20&0.65$\pm$0.18&0.38$\pm$0.07&0.36$\pm$0.06&5.66$\pm$0.81&-0.57$\pm$0.57\\

\hline
\end{tabular}
\end{center}
\end{table*}

\begin{table*}
\begin{center}
\caption{Pearson's correlation ($p_w$) and regression statistics ($R^2$) for M51a's ENRs PAH ratios and physical properties.}
\label{apptab:M51fluxPropertiesRandp}
\setlength{\tabcolsep}{15pt} 
\begin{tabular}{lccccccc}
\hline \\       
Physical Properties& Regions & \multicolumn{2}{c}{6.2/11.2} & \multicolumn{2}{c}{7.7/11.2} & \multicolumn{2}{c}{8.6/11.2} \\
& & $p_w$ & $R^2$ & $p_w$ & $R^2$ & $p_w$ & $R^2$ \\
\hline \\

$R_g/R_{25}$& ENRs & 0.74 & 0.54 & 0.82 & 0.63 & 0.67 & 0.44 \\
 & SF$_{ENR}$ & 0.32 & 0.11 & 0.27 & 0.07 & 0.37 & 0.12 \\
 & ISM$_{ENR}$ & 0.53 & 0.29 & 0.73 & 0.42 & 0.44 & 0.19 \\
$[{\rm S}\,\textsc{iv}]/[{\rm Ne}\,\textsc{ii}]$ & ENRs & -0.82 & 0.69 & -0.57 & 0.34 & -0.63 & 0.43 \\
  & SF$_{ENR}$  & -0.83 & 0.75 & -0.63 & 0.44 & -0.63 & 0.42 \\
  & ISM$_{ENR}$ & -0.37 & 0.05 & 0.13 & -0.004 & -0.52 & 0.24 \\
{$12 + \mathrm{log(O/H)}$ (dex)} & SF$_{ENR}$  & -0.79 & 0.60 & -0.30 & 0.05 & -0.66 & 0.43 \\
{T$_e${[}N\,\textsc{ii}{]} (K)} & SF$_{ENR}$  & 0.53 & 0.28 & 0.60 & 0.32 & 0.66 & 0.43 \\

\hline

\end{tabular}
\end{center}
\end{table*}

\begin{table*}

\begin{center}
\caption {Slopes ($\alpha$) and intercepts ($\beta$) of linear regression fits among PAH intensity ratios and physical parameters in M51a's ENRs.} 
\label{apptab:phys_prop_slopes}

\begin{tabular}{lccccccc}
\hline \\ 
Physical Properties & Regions & \multicolumn{2}{c}{6.2/11.2} & \multicolumn{2}{c}{7.7/11.2} & \multicolumn{2}{c}{8.6/11.2} \\
       &    & $\alpha$ & $\beta$ & $\alpha$ & $\beta$ & $\alpha$ & $\beta$ \\
\hline \\
$R_g/R_{25}$&ENRs&0.71$\pm$0.22&0.98$\pm$0.10&2.29$\pm$0.72&2.89$\pm$0.31&0.31$\pm$0.12&0.70$\pm$0.05\\
&SF$_{ENR}$&0.26$\pm$0.36&1.38$\pm$0.16&0.80$\pm$1.49&4.22$\pm$0.67&0.09$\pm$0.12&0.89$\pm$0.05\\
&ISM$_{ENR}$&0.46$\pm$0.23&0.96$\pm$0.09&1.50$\pm$0.72&2.77$\pm$0.29&0.20$\pm$0.17&0.70$\pm$0.07\\
$[{\rm S}\,\textsc{iv}]/[{\rm Ne}\,\textsc{ii}]$&ENRs&-3.15$\pm$0.65&1.97$\pm$0.08&-4.84$\pm$2.55&5.38$\pm$0.29&-0.67$\pm$0.25&1.03$\pm$0.03\\
&SF$_{ENR}$&-3.82$\pm$0.73&1.99$\pm$0.07&-6.48$\pm$3.16&5.57$\pm$0.29&-0.75$\pm$0.29&1.04$\pm$0.02\\
&ISM$_{ENR}$&-1.48$\pm$2.10&1.49$\pm$0.60&0.24$\pm$2.90&3.67$\pm$0.84&-0.75$\pm$0.71&1.05$\pm$0.21\\
$12 + \mathrm{log(O/H)}$ (dex)&SF$_{ENR}$&-0.99$\pm$0.29&10.15$\pm$2.50&-1.15$\pm$2.27&14.52$\pm$19.74&-0.32$\pm$0.13&3.71$\pm$1.16\\
T$_e${[}N\,\textsc{ii}{]} (K)&SF$_{ENR}$&2.18$\pm$1.46$^a$&0.10$\pm$0.94&12.6$\pm$4.97$^a$&-3.47$\pm$3.16&1.08$\pm$0.48$^a$&0.24$\pm$0.31\\
\hline
\end{tabular}
\\
$^a$ Values are multiplied by a factor of $10^{-4}$
\end{center}
\end{table*}

\begin{table*}
\begin{center}
\caption {Slopes ($\alpha$) and intercepts ($\beta$) of linear regression fits for PAH/VSG correlations among different regions and sources.} 
\label{Apptab:M51slopePAHVSG}
   
\begin{tabular}{@{}lccccccccc}
\hline

Regions & \multicolumn{2}{c}{log([PAH]/[VSG]) }&&&&&&\\

&\multicolumn{2}{c}{vs}&\multicolumn{2}{c}{PAH/VSG vs 6.2/11.2}&\multicolumn{2}{c}{PAH/VSG vs 7.7/11.2}&\multicolumn{2}{c}{PAH/VSG vs 8.6/11.2}\\
&\multicolumn{2}{c}{log([S\,\textsc{iv}]/[Ne\,\textsc{ii}])}&&&&&&\\

&$\alpha$&$\beta$&$\alpha$&$\beta$&$\alpha$&$\beta$&$\alpha$&$\beta$\\

\hline
NR&\ldots&\ldots&0.29$\pm$0.02&0.02$\pm$0.001&0.08$\pm$0.01&0.001$\pm$0.0001&0.35$\pm$0.05&0.07$\pm$0.009\\
ENRs 1-11$^a$&\ldots&\ldots&0.10$\pm$0.02&0.26$\pm$0.06&0.02$\pm$0.01&0.33$\pm$0.26&0.01$\pm$0.05&0.37$\pm$1.44\\
ENRs 1-11$^b$&\ldots&\ldots&0.07$\pm$0.04&0.31$\pm$0.17&0.03$\pm$0.13&0.38$\pm$1.53&-0.15$\pm$0.07&0.52$\pm$0.26\\
ENRs 1-2, 9-11&\ldots&\ldots&0.13$\pm$0.02&0.19$\pm$0.02&0.03$\pm$0.01&0.23$\pm$0.05&0.05$\pm$0.05&0.31$\pm$0.31\\
M51a$^a$&\ldots&\ldots&0.16$\pm$0.02&0.17$\pm$0.06&0.04$\pm$0.01&0.21$\pm$0.05&0.20$\pm$0.03&0.20$\pm$0.03\\
M51a$^b$&\ldots&\ldots&0.14$\pm$0.03&0.19$\pm$0.03&0.07$\pm$0.10&0.19$\pm$0.28&0.12$\pm$0.05&0.26$\pm$0.10\\
M33&-0.25$\pm$0.09&-0.16$\pm$0.06&1.42$\pm$0.34&-0.20$\pm$-0.05&0.43$\pm$0.09&-0.19$\pm$0.04&2.54$\pm$0.88&-0.47$\pm$0.16\\
M83&-0.35$\pm$0.09&-0.28$\pm$0.06&\ldots&\ldots&\ldots&\ldots&\ldots&\ldots\\
\hline
\end{tabular}
\\
$^a$ Determined without including outlier point with extremely high PAH/VSG.\\
$^b$ Determined with including outlier point with extremely high PAH/VSG.
\end{center}
\end{table*}

\end{document}